# A Systematic Mapping Study on Microservices Architecture in DevOps

**Muhammad Waseem [a], Peng Liang [a, *], Mojtaba Shahin [b]**

[a] School of Computer Science, Wuhan University, 430072 Wuhan, China
[b] Faculty of Information Technology, Monash University, 3800 Melbourne, Australia

m.waseem@whu.edu.cn, liangp@whu.edu.cn, mojtaba.shahin@monash.edu

**ABSTRACT**
**Context**: Applying Microservices Architecture (MSA) in DevOps has received significant attention in recent years. However, there exists no comprehensive review of the state of research on this topic.
**Objective**: This work aims to systematically identify, analyze, and classify the literature on MSA in DevOps.
**Method**: A Systematic Mapping Study (SMS) has been conducted on the literature published between January 2009 and July 2018.
**Results**: Forty-seven studies were finally selected and the key results are: (1) Three themes on the research on MSA in DevOps are "microservices development and operations in DevOps", "approaches and tool support for MSA based systems in DevOps", and "MSA migration experiences in DevOps". (2) 24 problems with their solutions regarding implementing MSA in DevOps are identified. (3) MSA is mainly described by using boxes and lines. (4) Most of the quality attributes are positively affected when employing MSA in DevOps. (5) 50 tools that support building MSA based systems in DevOps are collected. (6) The combination of MSA and DevOps has been applied in a wide range of application domains.
**Conclusions**: The results and findings will benefit researchers and practitioners to conduct further research and bring more dedicated solutions for the issues of MSA in DevOps.

# 1. Introduction

Microservices Architecture (MSA) is a cloud-native architectural style, which is inspired by Service-Oriented Architecture (SOA). Typically, microservices are organized as a suite of small granular services that can be implemented (developed, tested, and deployed) on different platforms through multiple technological stacks [1]. Each service of the MSA runs on its own process and communicate with each other through, e.g., RESTful or RPC-based APIs [2].

MSA has become popular in industry because of its benefits, such as availability, flexibility, scalability, loose coupling, and high velocity [3] According to the International Data Corporation (IDC), by the end of 2021, 80% of cloud-based applications will be developed using by MSA [1]. It is also argued that the worldwide DevOps market would grow to $5.6 billion in 2021 [4]. Another published report reveals that organizations may adopt MSA for different purposes [6], for example, to gain agility (82%), to improve organization performance (57%), and scalability (78%). This report also shows that the motivation behind implementing MSA in 47% of organizations was DevOps [5].

DevOps is a set of practices for developing, testing, and deploying software quickly and reliably by promoting collaboration between the developers, testers, and operators [6]. DevOps practices aim "*to decrease the time between changing a system and transforming that change*

* Corresponding author at: School of Computer Science, Wuhan University, China. Tel: +86 27 68776137; fax: +86 27 68776027. E-mail address: liangp@whu.edu.cn (P. Liang)

*into production environment*" [2]. Many practitioners and researchers advocate that MSA has a natural progression of embracing DevOps [7, 8]. DevOps brings additional productivity with MSA through using tools chain and a fast feedback mechanism [9].

To understand how MSA is employed in DevOps, we conducted an SMS through a collection of primary studies on MSA in DevOps context. The objective of this SMS is to *identify, analyze, and classify the literature on MSA in DevOps with respect to the research themes, problems, solutions, challenges, description methods, patterns, quality attributes (QAs), tools, and application domains*. The objective of this SMS is further decomposed into a number of Research Questions (RQs) that are listed in Table 1.

The key contributions of this SMS are: (1) A classification of the research themes related to MSA in DevOps. (2) A classification of the problems that practitioners may face during the implementation of MSA in DevOps and the solutions adopt to address the problems. (3) A list of identified research challenges in the context of MSA in DevOps. (4) A classification of the tools that support MSA in DevOps. (5) A list of MSA description methods, MSA patterns, QAs, tools, and application domains.

The rest of this paper is organized as follows: Section 2 briefly introduces MSA and DevOps, existing literature reviews, and motivation of this SMS. Section 3 presents the research method used in this study. Section 4 provides the study results. Section 5 discusses the results. Section 6 describes the threats to validity and Section 8 concludes the study.

# 2. Background

In this section, we provide an overview of MSA and DevOps, existing literature reviews, and motivation of this SMS.

## 2.1. Microservices Architecture

MSA style is gaining momentum for the development and deployment of software applications as a suite of small granular services that can be integrated through lightweight communication mechanisms, normally RESTful APIs [10]. Microservices are small, understandable components that hold the business capabilities around the services [11]. These services can be scaled independently (as they are loosely coupled) by implementing different technology stacks [2]. Many researchers and practitioners argue that MSA is an evolution of Services Oriented Architecture (SOA), as seen in the context of independent/self-management of services, and lightweight nature [12]. On the other hand, MSA can be differentiated from SOA in terms of component sharing, service communication, service mediation, and remote service access [13]. SOA is built on the idea of sharing as much as possible whereas MSA is formed on the idea of sharing as little as possible [13, 14]. MSA uses choreography style for inter-service communication whereas SOA employs orchestration style for service coordination. For service mediation, MSA uses the API layer that acts as the service facade while SOA adopts the concept of messaging middleware for service coordination. Moreover, MSA mostly relies on Representational State Transfer (REST) protocol and simple messaging as remote service access protocols; however, SOA can handle different types of remote access protocols including simple messaging for accessing remote services [13].

## 2.2. DevOps

DevOps is an approach based on agile and lean software development principals [15]. This approach promotes the collaboration of development and operations staff in order to develop quality software in a continuous manner [16]. DevOps consists of a set of practices (i.e., continuous planning, integration, deployment, testing, and monitoring) intended to develop, test, and deploy software changes quickly and reliably by promoting strong collaboration between the developers, testers, and operators [17]. Furthermore, DevOps automates the software deployment process from source code in the development environment to the production environment by utilizing different kinds of tools (see Section 4.4). This approach shortens the time to market by adopting continuous feedback, Continuous Integration (CI), and Continuous Delivery (CD) practices [9].

## 2.3. Motivation for this Mapping Study

DevOps and MSA are emerging trends in both industry and academia. Google Trends reveals that searches related to the terms "microservice" and "DevOps" are on the top in technology trends, and have been growing at an equal rate after 2014 [2]. MSA is the first architectural style after the DevOps revolution that has emerged, evolved, and widely adopted by the industry [18]. MSA based applications consist of 10s, 100s, or even 1000s of services that could be developed, tested, and deployed independently. All of these services and the infrastructures where the services are developed, tested, and deployed require robust automation to handle the number of the processes and velocity of change [19]. It is argued that DevOps can reduce the impact of the challenges related to MSA development and operations [20]. These challenges may occur due to distributed applications, poorly managed code infrastructure, lack of test case automation [20, 21], and no close alignment between development and operations activities [19].

DevOps is a culture that combines new or improved practices, processes, team structures and responsibilities, and tools to maximize the ability of an organization to deliver applications and services quickly [15, 22]. DevOps acts as a process framework that can be used for developing, deploying, and managing MSA [1]. The coexistence of microservices and DevOps enables reusability, decentralized data governance, automation, and built-in scalability [2]. MSA and DevOps have many common characteristics that make them a perfect fit for each other. For instance, DevOps practices and MSA promote the idea of decomposing large problems into smaller pieces and then address them through small cross-functional teams [23]. Containerized microservices can be realized independently because DevOps gives them a favor of continuous integration and deployment. Although it is not compulsory to design software systems based on MSA in DevOps, most of the challenges arisen in DevOps can be resolved by using MSA [17]. This combination is expected to increase the team's throughput and the overall quality of the system [1, 23]. For example, with MSA and DevOps, Netflix and Amazon engineers can do hundreds of deployments each day [19]. The MSA and DevOps combination brings several other benefits, including frequent software release, reliability and scalability of systems, resilience in the case of failure, and management of decentralized teams to control the application development [24, 25]. Moreover, the DevOps toolchain helps to continually code, build, test, package, release, configure, and monitor the MSA based systems. Furthermore, both MSA and DevOps are designed to offer great agility and operational efficiency for an enterprise [17].

Due to the growing importance of MSA in DevOps, an increasing amount of literature has been published through diverse venues in the last few years. Currently, many aspects are still unclear and scattered in literature in the context of MSA and DevOps combination. For instance, research themes, problems that prevent adopting MSA in DevOps, solutions to address those problems, and open research challenges. We believe that analyzing MSA in DevOps context may help practitioners to adopt this combination smoothly and will also help researchers to identify further research opportunities.

# 3. Research Methodology

The goal of this study is to get an overview of MSA in DevOps. More specifically, this study aims to identify problems, solutions, challenges, MSA description methods, patterns, qualities attributes, tools, application domains, and research opportunities in the context of MSA in DevOps. To that end, we conducted an SMS to collect, classify, and analyze the primary studies on MSA in DevOps. Another form of secondary research that can be used for conducting a review on a phenomenon is SLR. An SLR provides "*a means of identifying, evaluating, and interpreting all available research relevant to a particular research question*" [26]. While SLR enables researchers to conduct an in-depth analysis of the literature on a research area, SMS aims at covering the breadth of a research area. Moreover, SMS provides a systematic and objective procedure for identifying and classifying what evidence is available in a specific research area [27]. We decided to conduct an SMS as the scope of our studied topic (i.e., MSA in DevOps) is broad, and it includes many topics (e.g., design, implementation, migration, techniques, and tools). However, we did not only perform a mapping but also synthesized the data by using thematic analysis method [28]

To conduct this SMS, we followed the guidelines proposed in [27], complemented with the strategies presented by Kitchenham et al. for SLRs [26]. Figure 1 shows the process of this SMS, which was executed in three steps: (i) planning the mapping study, (ii) collecting and analyzing the data, and (iii) mapping and documenting the results.

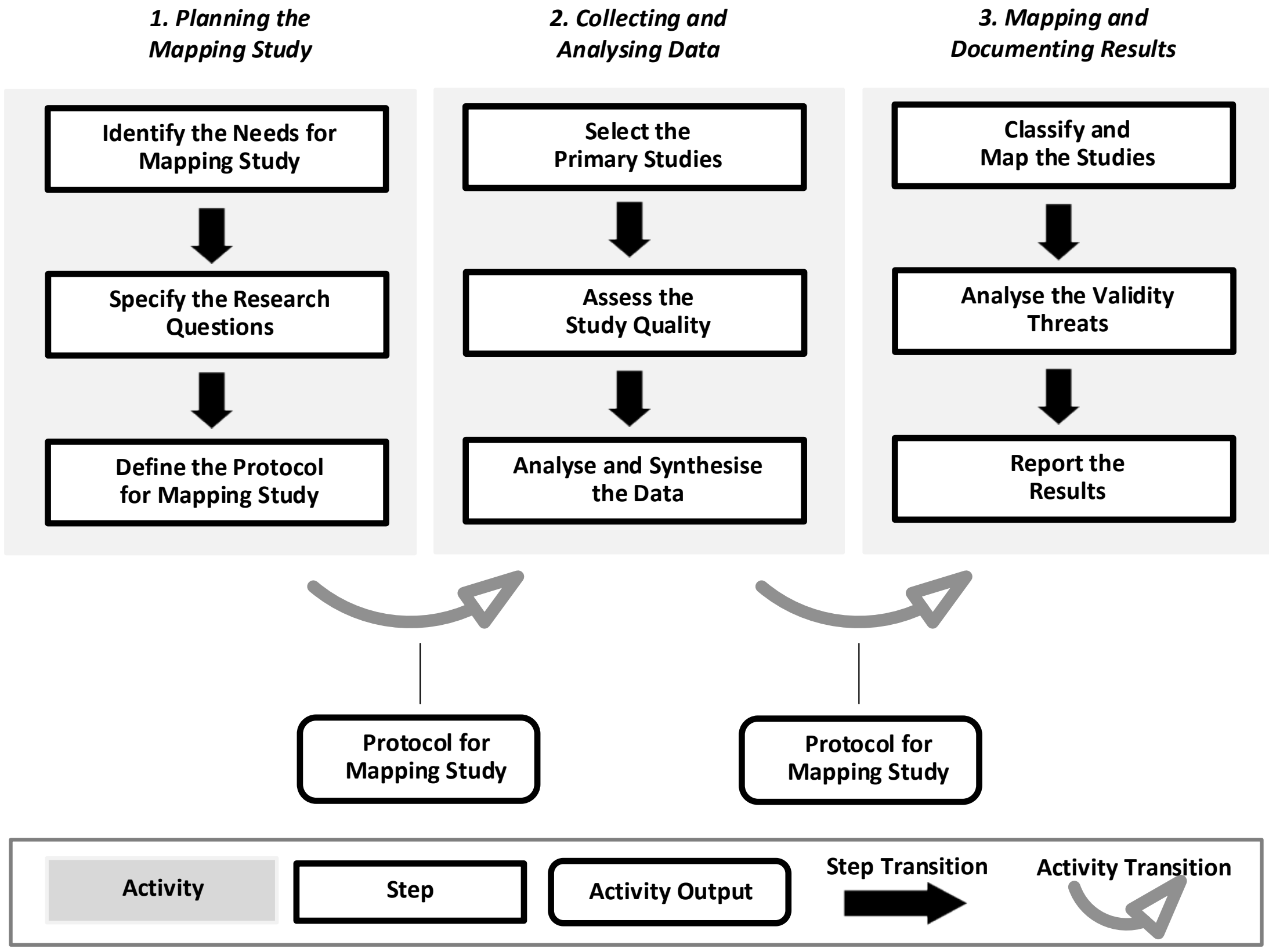

Figure 1. Process of this Systematic Mapping Study

## 3.1. Research Questions

To conduct this SMS, we derived the following eight RQs (as listed in Table 1) according to the goal of this study.

Table 1. Research Questions and their Rationale

| # | Research Question | Rationale |
|---|---|---|
| **Category 1: Demography, Classification, and Mapping of Research** | | |
| **#** | **Research Question** | **Rationale** |
| **RQ1.1** | What is the frequency and type of published research on MSA in DevOps? | This RQ aims to collect the data about intensity and type of publications on MSA in DevOps. The answer to this RQ will provide information about publication trends and prominent venues for MSA in DevOps research. |
| **RQ1.2** | What are the existing research themes on MSA in DevOps and how can they be classified and mapped? | The answer to this RQ will establish the foundation for systematic analysis of the existing research on MSA in DevOps through a taxonomy of the research themes. This taxonomy will provide (i) a base for classifying the existing research on MSA in DevOps and (ii) analyzing the current state of the art on MSA in DevOps context. |
| **Category 2: Problems, Solutions, and Challenges** | | |
| **#** | **Research Question** | **Rationale** |
| RQ2.1 | What problems have been reported when implementing MSA in DevOps? | The implementation of MSA in DevOps context is not without obstacles. The answer to this RQ will identify |
| | | and classify the problems related to the adoption of MSA in DevOps. |
| RQ2.2 | What solutions have been employed to address the problems? | A solution could be a best practice, a tool, a technique, or a framework. The answer to this RQ will help to identify the solutions to overcome the problems related to the implementation of MSA in DevOps context. |
| RQ2.3 | What challenges have been reported when implementing MSA in DevOps? | There might be open challenges that are reported in the selected studies without any proposed solutions. The answer to this RQ will identify and report them as future research opportunities in this area. |
| **Category 3: MSA Description Methods, Patterns, and Quality Attributes** | | |
| **#** | **Research Question** | **Rationale** |
| **RQ3.1** | What methods are used to describe MSA in DevOps? | MSA based applications can be designed and modeled by using different architecture description methods. The answer to this RQ will provide the information regarding MSA description methods (graphical, textual, or both) that have been used for expressing, communicating, and analyzing the features of MSA based systems in DevOps context. |
| **RQ3.2** | What MSA design patterns are used in DevOps? | Several design patterns are proposed to address the issues related to the implementation of MSA. The answer to this RQ will help to identify the MSA design patterns that are used to solve the common problems in DevOps context. |
| **RQ3.3** | What quality attributes are affected when employing MSA in DevOps? | QAs help to establish a static organization (e.g., modularity, testability, and maintainability, etc.) and dynamic behavior (e.g., throughput, robustness, and scalability, etc.) of a software application. The answer to this RQ will help to determine a set of QAs that have been positively or negatively affected when using MSA in DevOps context. |
| **Category 4: Tool support and Application Domains** | | |
| # | **Research Question** | **Rationale** |
| **RQ4.1** | What tools are available to support MSA in DevOps? | DevOps practices heavily rely on tool support and process automation [29]. The answer to this RQ will help to identify the tools that support or enable practicing MSA in DevOps. |
| **RQ4.2** | What are the application domains that employ MSA in DevOps? | The combination of MSA and DevOps has been frequently practiced in various application domains, such as entertainment, transportation, and E-commerce. The answer to this RQ will help to understand the application domains where the implementation of MSA in DevOps got much attention. |

## 3.2. Search Strategy

The search process for this study is divided into two phases. In Phase 1, we performed primary search by applying two search strings on the selected databases. Phase 2 employs the snowballing technique to complement Phase 1.

### 3.2.1. Phase 1: Primary Search

The primary search was based on querying digital databases (see Table 2) using customized search strings. We executed two search strings in parallel on the seven databases as shown in Figure 2. We limited our search to the peer-reviewed studies from January 2009 to July 2018. The year 2009 was chosen as the initial year because the term DevOps was introduced in 2009, but we did not find any relevant study until 2015. During the primary search, we covered the titles, abstracts, and keywords which resulted in a slightly high number of studies that were not relevant. We evaluated the primary studies through generic, specific, and quality assessment screening questions which are explained in Section 3.4 and summarized in Table 3 and Figure 2 After the primary search, the number of resulting primary studies reached to 45.

#### 3.2.2. Phase 2: Snowballing

In Phase 2, we inspected the references of the primary studies through the snowballing techniques [30] for further identification of the relevant studies. To maximize the chance of getting more studies, we performed the forward (i.e., collecting those studies citing the selected studies) and backward (i.e., using the references of the selected studies) snowballing to identify the relevant studies.

### 3.3. Search Strings

First, we created search String 1 based on the guidelines provided in [26]. To identify and select the primary studies, we formed search strings by considering the following factors:

- The research objective and RQs.
- Writing styles of microservices architecture (e.g., microservice, micro-services).
- Synonyms for "architecture" such as design and structure.
- No synonyms for DevOps.
- The limitations of the search engines in the digital databases.

During the pilot searches with search String 1, we observed that some well-known studies were not retrieved. We realized that these missing studies could be retrieved through a search string which only combines the two terms *microservice* and *DevOps*. Search String 2 was then formulated as (*microservice AND DevOps*), and we decided to use both search String 1 and search String 2 to retrieve the relevant studies. It is worth noting that the study selection procedure (i.e., see Section 3.4) was performed twice, one for each search string. Table 2 presents the search strings and databases used for searching the primary studies.

Table 2. Search Strings and Databases Used in this Systematic Mapping Study

| **Search strings** | | |
|---|---|---|
| **String 1**: ((microservi* OR micro-servi*) AND (architect* OR design OR structur*) AND DevOps)<br>**String 2**: (microservice AND DevOps) | | |
| **Databases** | | |
| **Database** | **Links** | **Targeted search area** |
| ACM Digital Library | http://dl.acm.org/ | Paper title, abstract |
| IEEE Explore | http://ieeexplore.ieee.org/ | Paper title, keywords, abstract |
| Springer Link | http://link.springer.com/ | Paper title, abstract |
| Science Direct | http://www.sciencedirect.com/ | Paper title, keywords, abstract |
| Wiley InterScience | http://onlinelibrary.wiley.com/ | Paper title, abstract |
| EI Compendex | https://www.engineeringvillage.com/ | Paper title, abstract |
| ISI Web of Science | https://login.webofknowledge.com | Paper title, keywords, abstract |

### 3.4. Screening and Qualitative Assessment

We followed the guidelines proposed in [31] to shortlist the retrieved papers from the databases. Our selection process includes screening studies and qualitatively assessing the selected studies.

#### 3.4.1. Screening Studies

Screening studies was conducted through a two-step process consisting of Generic Screening and Specific Screening. The former one includes six questions, i.e., GS1 to GS6 and the latter one has one question, i.e., SS1 (see Table 3). During the generic screening, the first author applied the basic criteria on the retrieved studies to ensure that there is no (i) duplicate studies, (ii) non-English papers, (iii) non-peer reviewed and white papers, (iv) secondary studies, (v) books, and (vi) studies published before 2009. During the specific screening, the first author also evaluated the selected studies with respect to available evidence about the problems, solutions, challenges, description methods, patterns, QAs, application domains, and available tools in the context of MSA and DevOps combination. The screening of studies was performed by the first author of this study through explicitly defining and following the criteria in Section Table 3. Then the second and third authors independently verified the screening results. All the researchers of this study have enough expertise, knowledge, and research experience about microservices and DevOps.

Table 3. Generic and Specific Screening Criteria for the Studies

| Code | Generic Screening Questions | Evaluation Scale | |
|---|---|---|---|
| GS1 | Is the study a duplicate study? | Yes | No |
| GS2 | Is the study written in English? | Yes | No |
| GS3 | Is the study peer-reviewed? | Yes | No |
| GS4 | Secondary study? | Yes | No |
| GS5 | Book? | Yes | No |
| GS6 | Is the study published between January 2009 to July 2018? | Yes | No |
| **Code** | **Specific Screening Questions** | **Evaluation Scale** | |
| SS1 | Does the study present problems, solutions, challenges, description methods, patterns, QAs, and tools in the context of MSA and DevOps combination? | Yes | No |

### 3.4.2. Qualitative Assessment of Studies

We took the guidelines for the qualitative evaluation from [32] and slightly tailored them according to our study. Moreover, we calculated the quality of the selected studies with Formula 1 [31]. This formula is based on the guidelines for the qualitative assessment of the selected studies [32]. The quality score was calculated through five generic (i.e., GI1 to GI5) and three specific assessment factors (i.e., SI1 to SI5) as listed in Table 4, in which each factor has a maximum score of 1. According to the criteria in [31], the contribution of the specific factors is more important than the generic factors. Consequently, we treated the accumulative value of S three times more than the accumulative value of G (i.e., 75% weight). To select the primary studies with decent quality, we decided to include the studies in this SMS that had an accumulative quality score greater than or equal to 1.5.

$$Quality\ Score = \left[\frac{\sum_{G=1}^{5}}{5} + \left(\frac{\sum_{S=1}^{5}}{5} \times 3\right)\right] \qquad (1)$$

Table 4. Qualitative Assessment Criteria for the Studies

| Code | Generic Items for Quality Assessment (Score: Yes 1, Partially 0.5, No 0) |
|---|---|
| GI1 | Are the problem definition and motivation of the study clearly presented? |
| GI2 | Is the research environment in which the study was conducted clearly explained? |
| GI3 | Is the research method used for the study clearly presented? |
| GI4 | Are the insights and lessons learned from the study explicitly mentioned? |
| GI5 | Are the limitations of the study explicitly discussed? |
| **Code** | **Specific Items for Quality Assessment (Score: Yes 1, Partially 0.5, No 0)** |
| SI1 | Does the study focus on MSA in DevOps? |
| SI2 | Does the study present problems, solutions, and challenges in the context of MSA and DevOps combination? |
| SI3 | Does the study present MSA description methods, MSA design patterns, QAs, and tools in the context of MSA and DevOps combination? |

## 3.5. Study Search

Figure 2 shows the search strings composition, database names, and the number of studies retrieved through each step of the study search process. We divided the search process into two phases: primary search and snowballing as described in Section 3.2.1 and Section 3.2.2 respectively.

**Step 1 - String execution**: We executed the two search strings separately on the seven databases (see Table 2) and retrieved 494 studies. Figure 2 explicitly shows the numbers of studies that we got by executing String 1 and String 2 on each database. For example, from ACM, we got 4 studies by executing String 1 and 47 studies by executing String 2.

**Step 2 - Study extraction**: We retrieved 494 studies by reading their titles and keywords. During this step, if we collected enough evidence to keep the paper for further reading, we marked that paper as "relevant" in our datasheet. If we were not sure to include or exclude the paper, we marked that paper as "doubted" and kept it for further reading. Otherwise, we marked that paper as "irrelevant." We selected 285 papers after the completion of this phase.

**Step 3 - Study screening**: The 285 studies were further assessed by reading their abstracts and conclusions. Each study was carefully inspected and ranked as "relevant," "irrelevant," or "doubted" according to the evidence available in the abstracts and conclusions. There were some studies where the first author of this SMS was unable to decide about inclusion or exclusion of the study under consideration. In such a situation, the first author transferred those studies to the second and third authors for their opinions about studies inclusion or exclusion. After completion of this step, we got 117 studies.

**Step 4 - Study selection**: In this phase, we read the full text of 117 studies and applied the inclusion and exclusion criteria on these selected papers. If a study met all the inclusion criteria (see Section 3.4), we included that study for this SMS. After completion of this phase, we got 45 studies.

**Snowballing**: We applied both forward and backward snowballing techniques on the 45 selected studies according to the guideline in [30]. In the beginning, we collected 451 titles through backward and forward (i.e., citations and references) snowballing. After the comparison with the titles of 494 retrieved studies, (which was the outcome of Step 1) and duplicate studies removal, we selected 86 studies for Step 2 (abstract and conclusion reading). We read the abstracts and conclusions of these 86 papers and picked 22 papers for full-text reading. By reading the full text of the papers and applying the generic, specific, and qualitative assessment screening criteria on 22 papers, we got only two studies (S46 and S47) through snowballing.

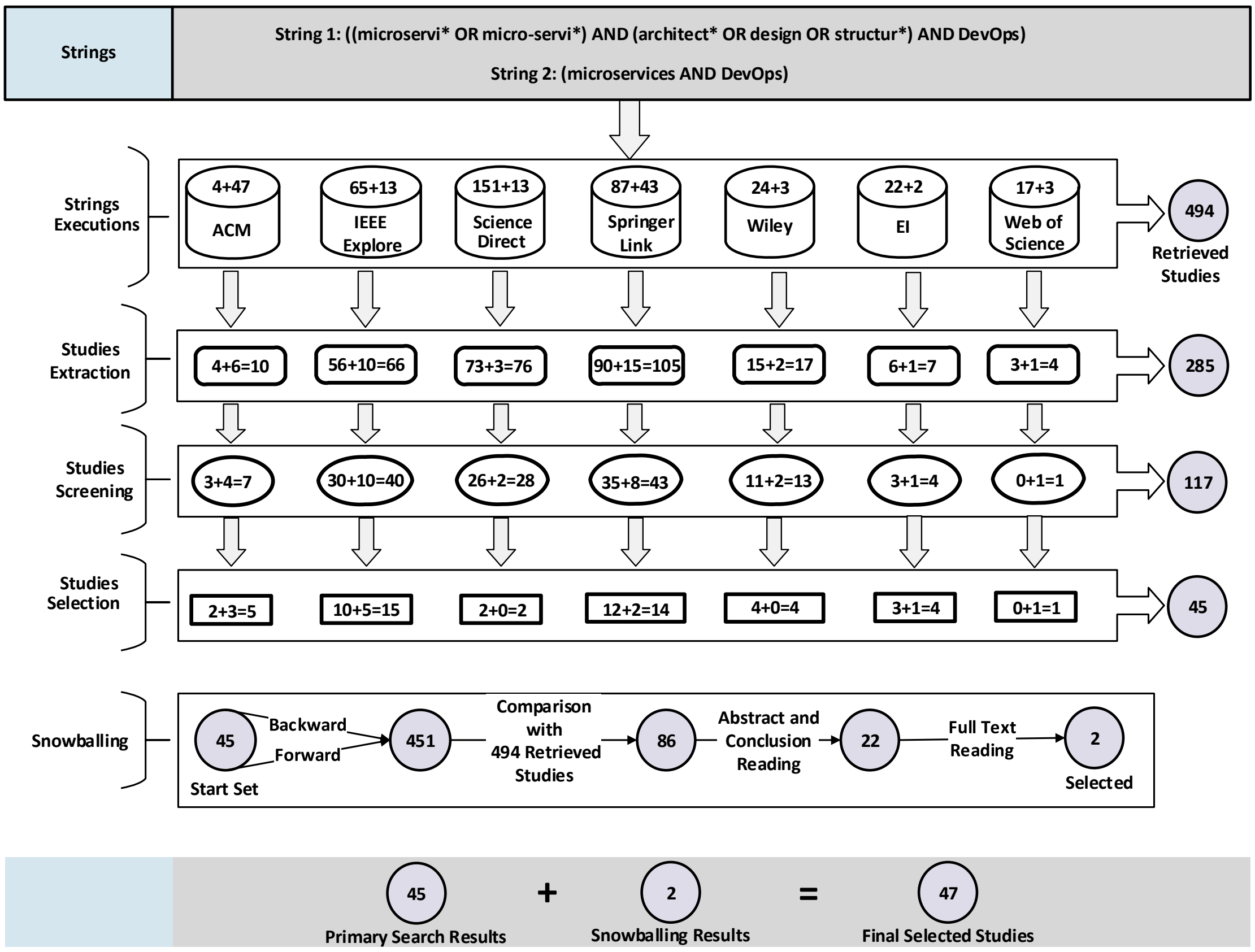

Figure 2. Steps and Results of the Study Search Process with Two Search Strings

## 3.6. Data Extraction and Synthesis

### 1) Data Extraction

For answering the RQs formulated in Section 3.1, we defined a set of data items (see Table 5) for extracting the required information from the selected studies. To check the reliability of the extracted data items, the first author conducted the pilot data extraction on ten studies, and the rest of the authors evaluated the extracted data. After the evaluation of the extracted data items, the first author used a revised set of data items for formal data extraction from the selected studies. All the authors then discussed the extracted data to reduce potential bias and ambiguity. Data items (D1-D3) are used to extract the general information of the selected studies, and the rest data items (D4-D15) are used to answer the RQs as shown in Table 1. The relationship between the data items and RQs is provided in Table 5. Finally, we used MS Excel sheets to record and further synthesize the extracted data.

Table 5. Data Items to be Extracted from the Selected Studies and Their Relevant RQs

| Code | Data Item | Description | Relevant RQ |
|---|---|---|---|
| D1 | Index | The ID of the study. | Overview |
| D2 | Study title | The title of the study. | Overview |
| D3 | Author(s) list | The full name of the authors. | Overview |
| D4 | Year | Publication year of the study. | RQ1.1 |
| D5 | Venue | The name of the publishing venue. | RQ1.1 |
| D6 | Publication type | Journal, conference, workshop, book chapter, and technical report. | RQ1.1 |
| D7 | Authors affiliation | Academia or industry or both. | RQ1.1 |
| D8 | Summary | The main idea of the study. | RQ1.2 |
| D9 | Problems | The problems reported in the study related to adoption of MSA in DevOps. | RQ2.1 |
| D10 | Solutions | The solutions proposed to solve the identified problems in the study. | RQ2.2 |
| D11 | Challenges | The problems reported without solutions in the study related to adoption of MSA in DevOps. | RQ2.3 |
| D12 | MSA description methods | The methods mentioned in the study for describing MSA design. | RQ3.1 |
| D13 | MSA patterns | The MSA patterns used in DevOps in the study. | RQ3.2 |
| D14 | Quality attributes | The QAs affected when employing MSA in DevOps. | RQ3.3 |
| D15 | Tool support | The tools mentioned in the study to support MSA in DevOps. | RQ4.1 |
| D16 | Application domains | The application domains that employ MSA in DevOps. | RQ4.2 |

**2) Data Synthesis**

We used descriptive statistics method for analyzing the data generated against the data items D1-D7, D12-D14, and D16. The qualitative data collected from data items D8 to D11, and D15 mostly comprise free text description (i.e., study' focuses, problems, solutions, and challenges). Therefore, we analyzed the qualitative data through thematic analysis [33] by following the steps: (1) Familiarizing with data: we repeatedly read the selected studies and noted down all points regarding the research purposes (D8), challenges (D9), solutions (D10), MSA description methods (D11), and tools (D14). (2) Generating initial codes: after data familiarization, we produced an initial list of codes from the extracted data about research purposes, problems, solutions, challenges, MSA description methods, and tools. (3) Searching for themes: during this step, we analyzed the initially generated codes and brought them under the broader level of themes. For instance, "Performance overhead due to fine-grain decomposition" (see Figure 6). (4) Reviewing themes: all the authors reviewed and refined the coding results with corresponding themes. During this step, we separated, merged, and dropped some themes based on mutual discussion between all authors. (5) Defining and naming themes: during this step, we defined and further refined all the themes under precise and clear names. To give an example, "Requirements of MSA based systems in DevOps" was named as a precise category in this step (see Figure 6).

# 4. Results

This section reports the results of the SMS after analyzing and synthesizing the extracted data from the selected papers. First, we report demography, classification, and mapping of the identified research themes in Section 4.1. Second, the problems, solutions, and challenges are reported in Section 4.2. We discuss the MSA description methods, patterns, and QAs in Section 4.3. Finally, tool support and application domains are reported in Section 4.4.

## 4.1. Demography, Classification, and Mapping of Research

### 4.1.1. RQ1.1: Publication Distribution

The distribution of publications per year is an integral part of secondary studies which provides information regarding published studies for each year on a specific topic. In general, the data analysis from this perspective indicates the interest of researchers and practitioners in a specific research area.

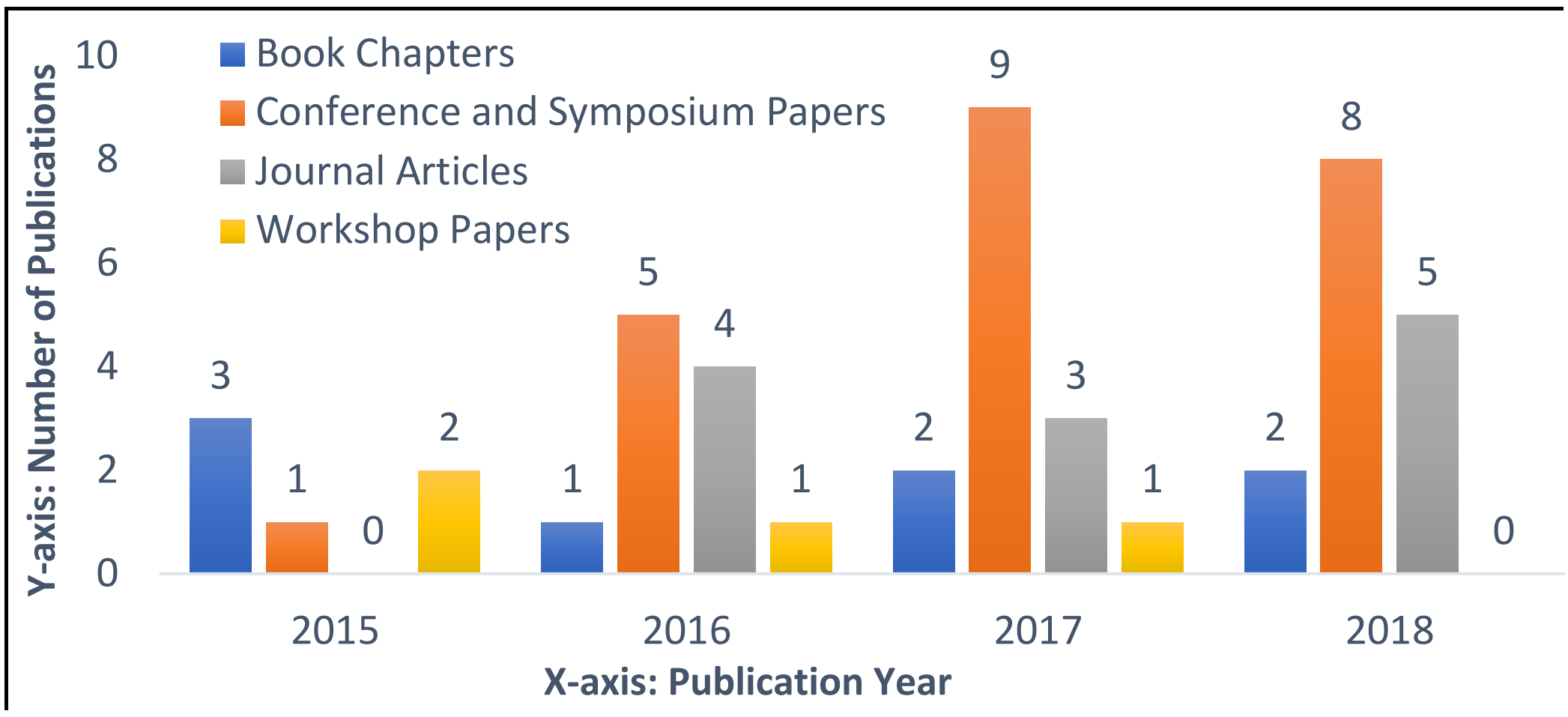

Figure 3. Studies Distribution over Publication Years

Figure 3 shows how publications (Y-axis) are distributed along the publication years (X-axis). Each year is mapped against the four colored bars, and each colored bar represents the study type published from 2015 to 2018. As shown in Figure 3, 5 studies were published in 2015, 11 studies were published in 2016, 15 studies were published in 2017, and 15 studies were published from January to July 2018. Note that we searched the studies from January 2009 to July 2018, however, studies included in this SMS were published between 2015 to July 2018.

Overall, we can see an upward trend in the number of the studies published per year especially from Jan 2016 to July 2018, and most of the studies (43 out of 47, 91.5%) had been published in that period, suggesting that researchers and practitioners are paying more attention to considering and adopting MSA in DevOps.

### 4.1.2. RQ1.1: Publication Type

Figure 4 shows that out of the 47 studies, 48.29% (i.e., 23 studies) were published in conference proceedings, and 25.53% (i.e., 12 studies) were published in journals, whereas this proportion for the book chapters and workshop papers is approximately 17.02% (i.e., 8 studies) and 8.51% (i.e., 9 studies) respectively.

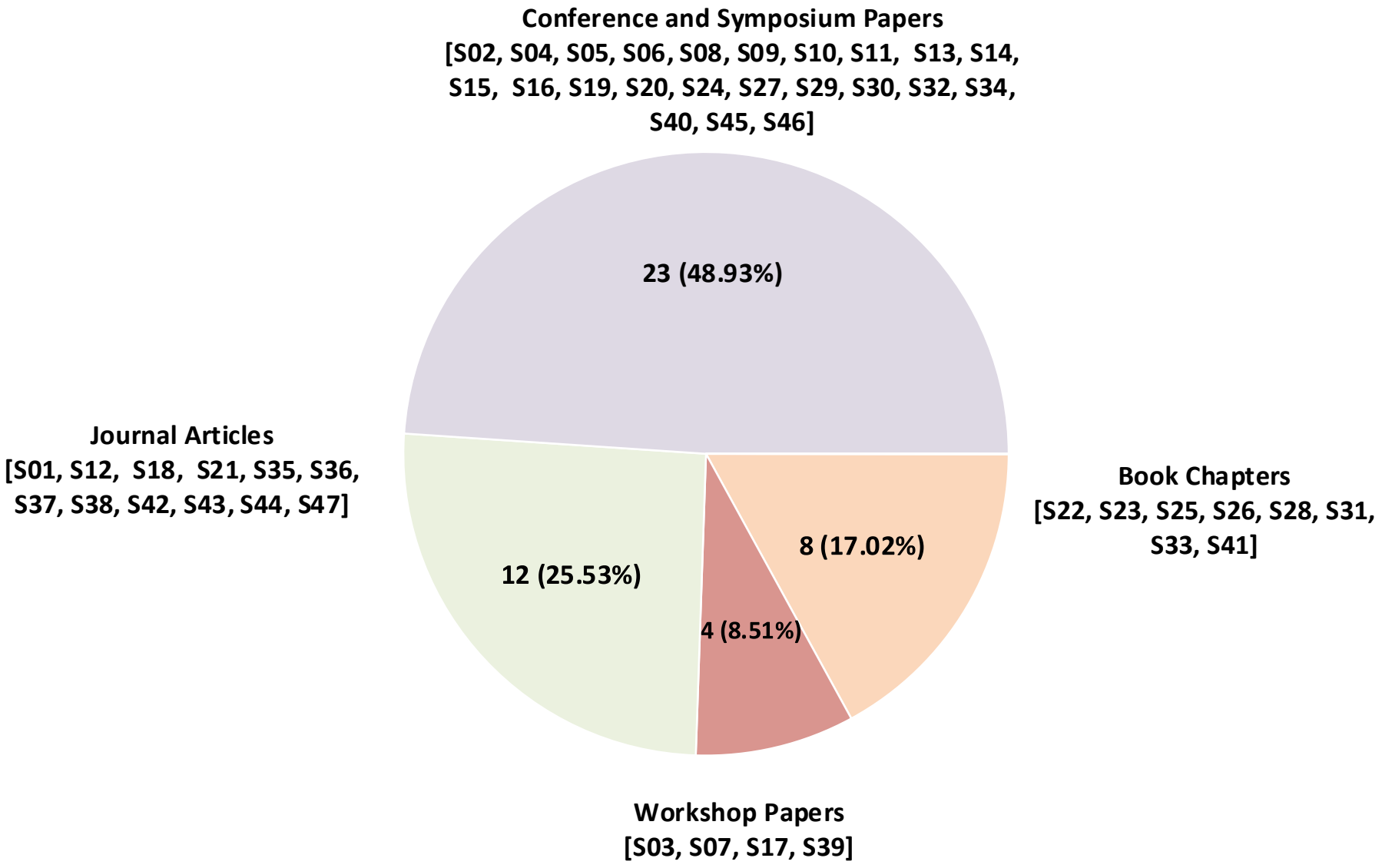

Figure 4. Overview of the Publication Type

### 4.1.3. RQ1.1: Publication Venues

The list of publication venues, types of publications, and the number and percentage of the selected studies published in each venue are summarized in Table 6. The 47 selected studies were published in 41 venues that are classified into four categories. It is noted that 80.48% (33 out of 41) of the venues belong to "Internet, Cloud, and Services Computing" and "Software Engineering" categories.

**Internet, Cloud, and Services Computing**: Journals and conferences that mainly invite the research outcomes related to internet, cloud, and service-oriented computing, such as cloud infrastructure, cloud applications, cloud management and operations, service-oriented software engineering, semantic services, service infrastructure, and development of service-oriented applications. This is the most popular category, which occupies 43.90% (18 out of 41) of the publication venues.

**Software Engineering**: Journals, books, and conferences that mainly publish the topics on software requirements, architecture, design, testing, maintenance and evolution, process improvement, project management, and software tools. In a broad sense, services computing is part of software engineering, but we treated it as a category since this SMS is service related. 39.02% (16 out of 41) of the publication venues belong to the software engineering category.

**Telecommunications and Networks**: Journals and conferences that mainly include the topics on telecommunication, information and communication technologies, network virtualization, 5th generation networks, internet of things, and network security and privacy. 9.75% (4 out of 41) of the publication venues are identified under this category.

**Multi-Disciplinary Computing**: Journals and publications that bring together trends, theories, practices, and experiences related to general computing disciplines. This category has 9.75% (4 out of 41) of the publication venues.

Table 6. Numbers and Proportions of the Selected Studies over the Venues

| Category | Publication Venue | Type | # | % |
|---|---|---|---|---|
| Interne, Cloud, and Services Computing | Microservices, IoT, and Azure | Book Chapter | 2 | 4.2 |
| | World Wide Web Journal | Journal | 1 | 2.1 |
| | IEEE Internet Computing | Journal | 1 | 2.1 |
| | IEEE Cloud Computing | Journal | 1 | 2.1 |
| | ACM Transactions on Internet Technology | Journal | 1 | 2.1 |
| | Software-Defined Cloud Centers: Operational and Management Technologies | Book Chapter | 1 | 2.1 |
| | Business in Real-Time Using Azure IoT and Cortana Intelligence Suite | Book Chapter | 1 | 2.1 |
| | International Conference on Ubiquitous Information Technologies and Applications (CUTE) | Conference | 1 | 2.1 |
| | European Conference on Service-Oriented and Cloud Computing (ESOCC) | Conference | 1 | 2.1 |
| | International Conference on Cloud Computing and Services Science (CLOSER) | Conference | 1 | 2.1 |
| | International Conference on Cloud Engineering (IC2E) | Conference | 1 | 2.1 |
| | IEEE Symposium on Service-Oriented System Engineering (SOSE) | Conference | 1 | 2.1 |
| | European Conference on Service-Oriented and Cloud Computing (ESOCC) | Conference | 1 | 2.1 |
| | International Conference on Service-Oriented Computing (ICSOC) | Conference | 1 | 2.1 |
| | International Conference on Exploring Services Science (IESS) | Conference | 1 | 2.1 |
| | International Conference on Services Computing (ISCC) | Conference | 1 | 2.1 |
| | International Conference on AI & Mobile Services (AIMS) | Conference | 1 | 2.1 |
| Software Engineeri | IEEE Software | Journal | 3 | 6.3 |
| | Journal of Software: Evolution and Process | Journal | 2 | 4.2 |
| | Microservices From Day One | Book Chapter | 1 | 2.1 |
| | Managing Software Crisis: A Smart Way to Enterprise Agility | Book Chapter | 1 | 2.1 |
| | The DevOps Adoption Playbook: A Guide to Adopting DevOps in a Multi-Speed IT Enterprise | Book Chapter | 1 | 2.1 |
| | Theory and Practice of Formal Methods | Book Chapter | 1 | 2.1 |
| | ACM/SPEC International Conference on Performance Engineering (ICPE) | Conference | 2 | 4.2 |
| | International Symposium on Software Engineering for Adaptive and Self-Managing Systems (SSASMS) | Conference | 1 | 2.1 |
| | International Conference on Software Architecture (ICSA) | Conference | 1 | 2.1 |
| | International Conference on Web Engineering (ICWE) | Conference | 1 | 2.1 |
| | International Conference on Software Engineering for Defence Applications (SEDA) | Conference | 1 | 2.1 |
| | International Conference on Information Systems Development (ISD) | Conference | 1 | 2.1 |
| | International Conference on AI & Mobile Services (AIMS) | Conference | 1 | 2.1 |
| | International Workshop on Quality-Aware DevOps (IWQAD) | Workshop | 1 | 2.1 |
| | Central Europe Workshop on Continuous Software Engineering (CSE) | Workshop | 1 | 2.1 |
| | International Enterprise Distributed Object Computing Workshop (EDOCW) | Workshop | 1 | 2.1 |
| **Telecommunications and Networks** | Transactions on Emerging Telecommunications Technologies | Journal | 1 | 2.1 |
| | International Conference on Information and Communication Technology Convergence (ICTC) | Conference | 2 | 4.2 |
| | International Conference on Telecommunications Network Strategy and Planning Symposium (NETWORKS) | Conference | 1 | 2.1 |
| | IEEE/IFIP Network Operations and Management Symposium (NOMS) | Conference | 1 | 2.1 |
| **Multi-Disciplinary Computing** | Procedia Computer Science | Journal | 2 | 4.2 |
| | Hawaii International Conference on System Sciences (HICSS) | Conference | 1 | 2.1 |
| | International Conferences on Practice and Experience on Advanced Research Computing (PEARC) | Conference | 1 | 2.1 |
| | International Conference on Current Trends in Theory and Practice of Informatics (SOFSEM) | Conference | 1 | 2.1 |

#### 4.1.4. RQ1.2: Mapping of Research Themes

Figure 5 and Table 7 show the classification of various research themes and subthemes that have been extracted from the selected studies by following the guidelines of thematic analysis in [33]. The result shows that the most discussed subthemes are *Approaches* and *Tools* which have been discussed in 13 and 12 studies respectively, whereas the least discussed subtheme is *Monitoring* of MSA based systems in DevOps with 4 studies. it is worth mentioning, several studies are classified into more than one themes or subthemes. For example, Study (S02) discussed the QAs concerns, testing, and monitoring of MSA based systems; Study (S07) discussed the architectural tactics/strategies for MSA in DevOps along with tool support.

**Thematic classification** organizes the selected studies based on the focus of research, and the selected studies are organized in three general themes: (1) Microservices development and operations in DevOps, (2) approaches and tool support for MSA based systems in DevOps, and (3) MSA migration experience in DevOps.

**Sub-thematic classification** provides a detailed view of the thematic classification, and eight subthemes are derived from the thematic classification: i) QAs concerns (6 studies), ii) design (11 studies), iii) development and deployment (12 studies), iv) testing (7 studies), v) monitoring (4 studies), vi) approaches (12 studies), vii) tools (13 studies), and viii) migration (7 studies). It is also worth mentioning that several studies discussed more than one subtheme. For example, Studies (S01) and (S09) discussed the design, and development and deployment subthemes.

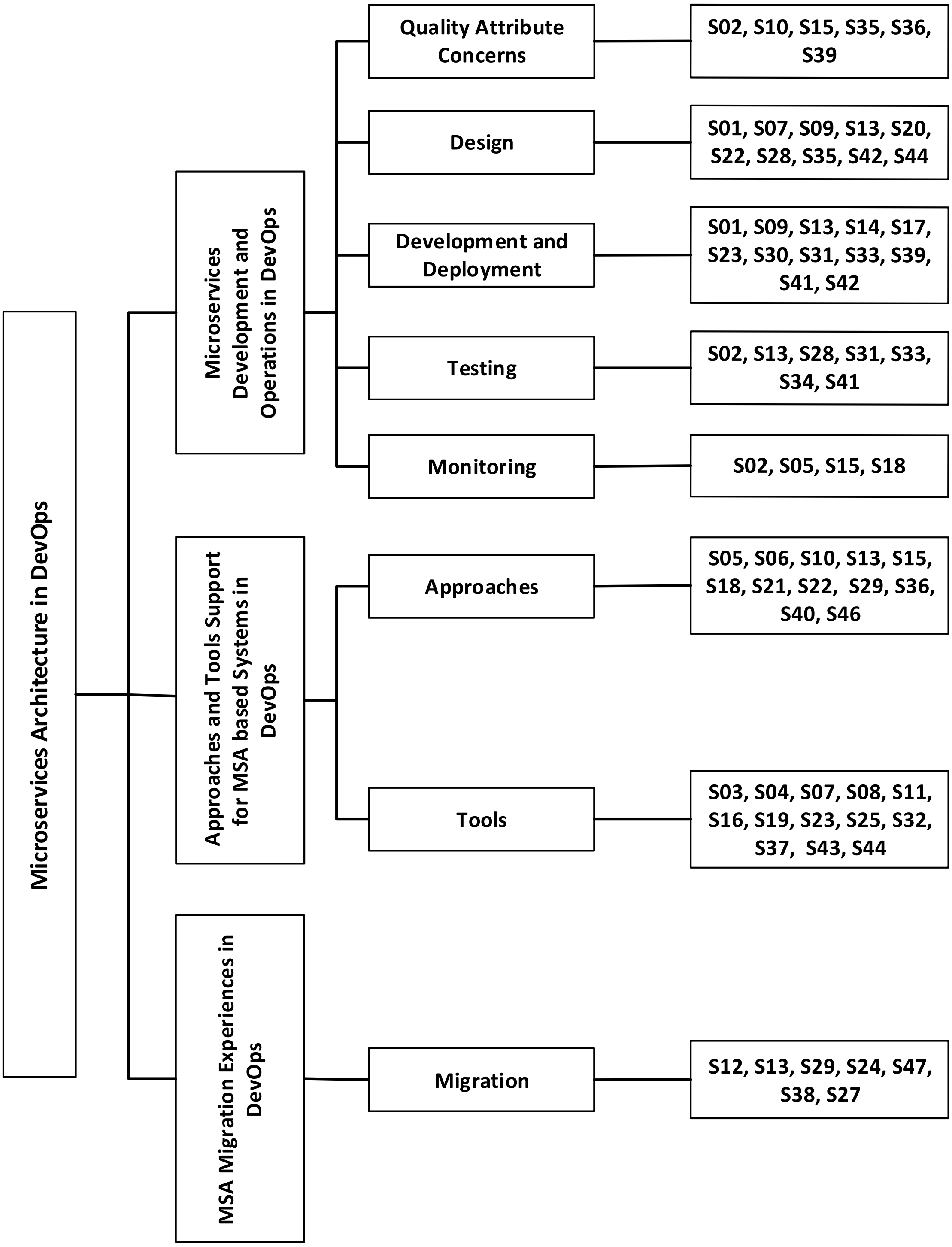

Figure 5. A Classification of Research Themes on MSA in DevOps from the Selected Studies

Table 7. Themes, Subthemes, and Key Points in the Classification of the Selected Studies

| Themes | Subthemes | Key Points with Their Study IDs |
|---|---|---|
| Microservices Development and Operations in DevOps | Quality Attribute Concerns | • Privacy and security concerns of microservices in DevOps (S10, S35, S36)<br>• Performance modeling of MSA based systems in DevOps (S02, S15)<br>• Non-functional requirements (e.g., scalability, exchangeability, reusability, code quality) for MSA based systems in DevOps (S39) |
| | Design | • MSA as an architectural pattern for cloud-native software systems (S01)<br>• Architectural tactics/strategies for MSA and DevOps in Big Data as Service (BDaaS) platforms (S07)<br>• MSA-driven design decisions about deployment and resource management in DevOps (S35)<br>• Microservices-driven design for cloud-based DevOps's infrastructure (S09, S42)<br>• Architecting for MSA based systems in CD and DevOps (S20)<br>• A container-based reference MSA for the deployment of OpenStack (S09)<br>• A reference architecture of the cloud's infrastructure that supports a combination of microservices and DevOps (S22, S44)<br>• Domain Driven Design (DDD) for microservices in DevOps (S13, S28) |
| | Development and Deployment | • Development and management of the microservices for cloud-native systems in DevOps (S01, S41, S42)<br>• Development and container-based deployment of microservices for OpenStack (S09)<br>• Development of MSA based mobile applications in DevOps (S13)<br>• Development and operations of MSA based enterprise systems (e.g., online store) in BizDevOps environment (S17)<br>• Development and deployment of MSA based enterprise systems (e.g., banks, customer relationship management) in DevOps (S33)<br>• Development, deployment, and packaging of microservices using Microsoft Azure in DevOps (S23)<br>• Development of MSA based eServices in DevOps (S30)<br>• Continuous deployment of microservices in DevOps (S31)<br>• Development and deployment of MSA based applications for the connected car in DevOps (S39)<br>• Automated deployment of microservices for SmartX Internet of Things (IoT) in DevOps (S14) |
| | Testing | • Performance testing of microservices in DevOps (S02)<br>• Testing of MSA based mobile applications in DevOps (S13)<br>• Testing strategies for microservices in CD (S28)<br>• A pre-production testing of microservices in DevOps (S31)<br>• Load testing of microservices in DevOps (S33)<br>• Integration testing of MSA based Smart-Energy IoT-Cloud Services in DevOps (S34)<br>• A/B testing for MSA based systems in DevOps (S41) |
| | Monitoring | • Performance monitoring of MSA based systems in DevOps (S02, S15)<br>• Monitoring of MSA based systems through a quality-aware DevOps approach (S05).<br>• Monitoring of microservices running inside the containers in DevOps (S18) |
| **Approaches and Tool support for MSA based systems in DevOps** | Approaches | • A quality-aware DevOps approach (i.e., Omina) for monitoring MSA based systems (S05)<br>• Autonomic Management System (AMS) for performance monitoring of MSA based systems in DevOps (S15)<br>• A microservices development and monitoring framework (i.e., Unicorn) with cloud-based DevOps IDEs support (S18)<br>• A Web-based deployment method for (micro)services that supports DevOps (S06)<br>• A DevOps framework (i.e., ARCADIA) that considers security and privacy requirements across development lifecycle of MSA based systems (S10)<br>• A migration approach for development and deployment of MSA based systems in DevOps (S13)<br>• A DevOps based Decision Support System (DSS) for the migration from monoliths to microservices (S29)<br>• A microservices deployment framework (i.e., SMART VM) that supports DevOps (S21)<br>• A Next-Gen DevOps solution for the multi-cloud environment that supports the development of MSA based systems (S22)<br>• An approach for protecting individual microservice and enabling autonomy of the DevOps teams (S36)<br>• A model-based approach (i.e., AUTOGENIC) for generating self-configuring microservices (S40)<br>• An incremental integration method for microservices that supports CD/CI/DevOps (S46) |
| | Tools | • A development and deployment platform that supports the combination of MSA and DevOps (S03, S04)<br>• A CAOPLE Integrated Development Environment (i.e., CIDE) for development of MSA based systems with DevOps support (S08, S11).<br>• A Big Data as a Service (BDaaS) platform that utilizes MSA and DevOps for fast development and delivery of the required services (S07, S16)<br>• An MSA based Network Function Virtualization (NFV) platform that supports DevOps for the development and deployment of MSA based systems envisioned for 5G networks (S19, S43)<br>• An MSA development, deployment, and testing platform (i.e., Microsoft Azure) that supports DevOps (S23)<br>• A tool (i.e., Jolie Redeployment Optimiser) for automatic and optimized (i.e., less expansive) deployment of microservices in DevOps (S25)<br>• An application (e.g., microservices) deployment scheduling tool based on multivendor environment workflow and DevOps practices (i.e., CD, CI) (S32)<br>• Automation tools that support DevOps (S37)<br>• Tool support for microservices' CD pipeline (S44) |
| **MSA Migration Experiences in DevOps** | Migration | • Migration of Mobile Back end as a Service (MBaaS) to MSA in DevOps (S12).<br>• Patterns to migrate from monoliths to MSA in DevOps (S12)<br>• Migrate monolithic mobile application to MSA based application in DevOps (S13)<br>• A DevOps based DSS for the migration from monoliths to microservices (S29)<br>• Motivations and challenges of migrating from monoliths to MSA in DevOps (S24, S47)<br>• DevOps practices that enable the migration of a monolithic architecture to MSA (S38)<br>• Migration of a monolithic architecture to cloud-native MSA in CD and DevOps (S27) |

## 4.2. Problems, Solutions, and Challenges

This section presents the problems, solutions, and challenges identified from the selected studies to answer RQ2 "*What problems, solutions, and challenges have been reported when implementing MSA in DevOps?*". By applying thematic analysis [33] on the extracted data for the classification of the identified problems and solutions, we derived 8 categories and 15 subcategories of themes. The eight problem and solution themes are (i) requirements of MSA based systems in DevOps, (ii) design of MSA based systems in DevOps, (iii) implementation of MSA based systems in DevOps, (iv) testing of MSA based systems in DevOps, (v) deployment of MSA based systems in DevOps, (vi) monitoring of MSA based systems in DevOps, (vii) organizational challenges, and (viii) resource management problems. Overall, we identified the solutions for 24 problems and provided the mapping between the problems and solutions Figure 6 shows the classification of problems and their corresponding solutions, whereas Table 8 presents the challenges which need further research. It should be noted that the problems that are reported without any solutions in the selected studies are referred to as research challenges. In other words, if a problem has at least one solution in any of the selected studies, we will not consider that problem as a challenge. We also identified some solutions that can address the problems in another category of problems. For example, as shown in Figure 6, the CIDE platform can address the problem of "Performance overhead due to fine-grain decomposition" in the category of "Requirements of MSA based systems in DevOps", and the problem of "Modification and integration of new functionality in existing microservices" in the category of "Implementation of MSA based systems in DevOps".

The classification shows that the problems related to two themes "Requirements of MSA based Systems in DevOps" and "Implementation of MSA based Systems in DevOps" are reported in more than half of the selected studies. We briefly explain the problems and solutions below.

### 4.2.1. RQ2.1 and RQ2.2: Problems and Solutions

**Requirements of MSA based systems in DevOps**: This category reports the problems and solutions related to the requirements of MSA based systems. To address performance overhead issues, Study (S08) presents DevOps based CAOPLE language Integrated Development Environment (CIDE). This platform provides precise control over the deployment and testing of microservices to address the performance overhead. Study (S33) proposes the VM auto-configuration methodology to address performance issues; VM auto-configuration method creates the central domain control agent for optimizing the performance of MSA based systems. Study (S18) proposes Unicorn framework to avoid delays and network performance issues, whereas Study (S24) suggests that architects should try not to decompose microservices too fine-grain. Study (S16) presents a DevOps based approach called "Neo-Metropolis". This approach offers open source solutions (e.g., Terraform, Ansible, Mesos, and Hadoop) to deal with scalability and elasticity of MSA based systems across different cloud platforms. Study (S18) argues for the use of containers to deal with scalability issues because containers provide an easy way to scale operations by creating more copies of the services (S18). Study (S41) suggests that developing microservices around business capabilities can address this scalability issue.

**Design of MSA based systems in DevOps**: This category reports the problems and solutions related to the design of MSA based systems in DevOps (see Figure 6), which can be further classified into application decomposition (S28, S33, S35, S37), security and privacy (S10, S18, S20, S36), and uncertainty (S01). Study (S28) recommends the Domain-Driven Design (DDD) pattern to address the application decomposition problem. By applying the DDD pattern, architects identify the bounded context (capabilities within the system) that can be used as a starting point for defining microservices. Similarly, Study (S33) recommends the Model-View-Controller (MVC) pattern for application decomposition into microservices in terms of business scope, functionalities, and responsibilities. Study (S10) presents the DevOps based ARCADIA framework to address security issues. This framework enables security and privacy across the microservices development lifecycle by providing multi-vendor security solutions (e.g., FWaaS and OAuth 2). Study (S18) presents DevOps based Unicorn framework that offers policies and constraints to meet security requirements of MSA based systems, whereas Study (S36) suggests that a combination of standard cryptographic primitives (e.g., hash and MAC functions for authentication encryption) can provide a high level of security to microservices communication and flexible authentication to DevOps teams. To deal with uncertainty issues in cloud-native architecture, Study (S01) proposes the theory-based control models at runtime patterns. Models at runtime patterns address the uncertainty aspects (e.g., resource availability) dynamically through the control loop.

**Implementation of MSA based systems in DevOps**: The identified problems and solutions in this category belong to microservices integration and managing databases for microservices. To deal with the operational and configuration complexity issue, Study (S20) recommends a CD platform, which provides a CD pipeline for each service that can give control over the integration of microservices. To address the complexity issues due to a large number of microservices, Study (S24) suggests two guidelines: first, keep the interface of each microservice as simple as possible for integration purposes, and second, it is recommended to use the technology which does not require specific programming language while implementing microservices to avoid from integration issues. Moreover, Study (S03) proposes a platform (i.e., HARNESS) to facilitate the integration of microservices that are developed in geographically distributed locations. In addition to these guidelines, Study (S08) also proposes the CIDE platform that provides precise control over testing, deployment, and integration of the new functionality into existing systems. To handle the problem of data management of MSA based systems, Study (S24) discusses the use of database per service and a shared database for multiple microservices patterns. Database per service pattern can be implemented through defining a separate set of tables per function, scheme per service, and database server per service, whereas a shared database pattern can be implemented by defining a single database for a group of microservices. Usually, microservices are grouped according to the business context to use the shared database.

**Testing of MSA based systems in DevOps**: The number of services, inter-communication processes, dependencies, instances, and other variables influence the testing process for MSA based systems in DevOps. We identified six studies that stress on excessive testing of MSA based systems in DevOps. Study (S28) claims that all traditional testing strategies (e.g., unit testing, functional testing, regression testing, etc.) can be used to test MSA based systems. Moreover, Study (S28) also recommends internal testing, service testing, protocol testing, composition testing, protocol testing, scalability/throughput testing, failover/fault tolerance testing, and penetration testing strategies. Apart from the testing strategies mentioned above, Study (S08) and Study (S11) presents the CIDE platform that can be used to test MSA based systems in DevOps. The tools we identified from the selected studies that can be used to test MSA based systems are listed in Figure 7.

**Deployment of MSA based systems in DevOps**: Many solutions have been proposed to address the issues of MSA based system deployment in DevOps (e.g., complexity, dynamic deployment, and deployment in development, production, and testing environment). For instance, Study (S12) recommends a multipurpose Docker Compose tool, which can work in different environments, such as staging, development, deployment, and testing environments, and smooth the deployment process of microservices in the development environment. Study (S27) recommends Kubernetes, working with a range of containers tools (e.g., Dockers), to deploy and scale microservices into the production environment. Study (S20) recommends that the frequent deployment of microservices must be automated through a CD pipeline to finish within due time. To address the problem of complexity in dynamic deployment of many microservices, Study (S08) and Study (S11) present the CIDE platform, which provides precise control over dynamic deployment through Communication Engine (CE) and Local Execution Engine (LEE). To deal with the problem of MSA based SaaS deployment, Study (S21) proposes the SmartVM framework to automate the deployment of MSA based SaaS. Study (S21) also provides strategies (e.g., Traefik, HTTP reverse proxy, round-robin) for load balancing and separating the functional and operational concerns. Jolie Redeployment Optimiser (JRO) has been employed to achieve an optimal deployment of MSA based systems (S25). JRO consists of three components: Zephyrus, Jolie Enterprise (JE), and Jolie Reconfiguration Coordinator (JRE), in which Zephyrus generates detailed and optimal architecture for MSA based systems, JE provides a framework for deploying and managing microservices, and JRE interacts with Zephyrus and JE for optimized deployment.

**Monitoring of MSA based Systems in DevOps**: A factory design pattern-based approach, called Omnia, has been proposed to address monitoring infrastructure problem (S05). This approach provides a component called monitoring interface, which enables developers to monitor MSA based systems independently and helps system administrators to build monitoring systems that are compatible with such interface by using monitoring factory components. Some tools can help address logging issues (see Figure 7). To address the problem of monitoring fine-grain microservices at runtime in a shared execution environment, Study (S18) presents DevOps based Unicorn framework, which can monitor highly decomposed MSA based systems at runtime (S18).

**Organizational Problems**: This theme reports problems related to culture, people, cost, and organization and team structure in the context of MSA and DevOps combination. To handle the problems that may be faced with when introducing MSA and DevOps combination in a given organization, Study (S23) suggests some guidelines, such as adopting new organizational structure, introducing small cross-functional teams, training for learning new skills (e.g., MSA, DevOps), changing employee habits toward the team work and sharing of responsibilities, and providing separate physical locations to teams, etc. Study (S24) suggests that the monolithic organizational structure needs to be aligned with the architecture of MSA based systems. Similarly, to address the issue related to establishing skilled and educated DevOps teams, Study (S24) suggests that the organization should arrange training programs for their employees for learning and adopting microservices in DevOps.

**Resource Management Problems**: This category provides the mapping of problems and solutions for different types of resources required to implement MSA in DevOps. Study (S01) recommends the virtualization of applications, infrastructures, and platforms resources as a solution for addressing resource management problems. Study (S09) suggests using containers and VMs for microservices in DevOps to get the desired level of efficiency in resource utilization. Study (S03) proposes the HARNESS approach (i.e., a DevOps based approach) that provides a cloud-based platform for bringing together commodity and specialized resources (e.g., skilled people). Study (S19) introduces an MSA based SONATA NFV platform with DevOps to address resource management problems by providing a set of tools (e.g., GitHub, Jenkins, Docker). The SONATA NFV platform can also create the CI/CD pipeline to automate steps in software delivery process. Study (S09) argued that dedicated access to the host's hardware can be increased either by giving extra privileges to microservices or by enhancing the capability of containers to access the host resources.

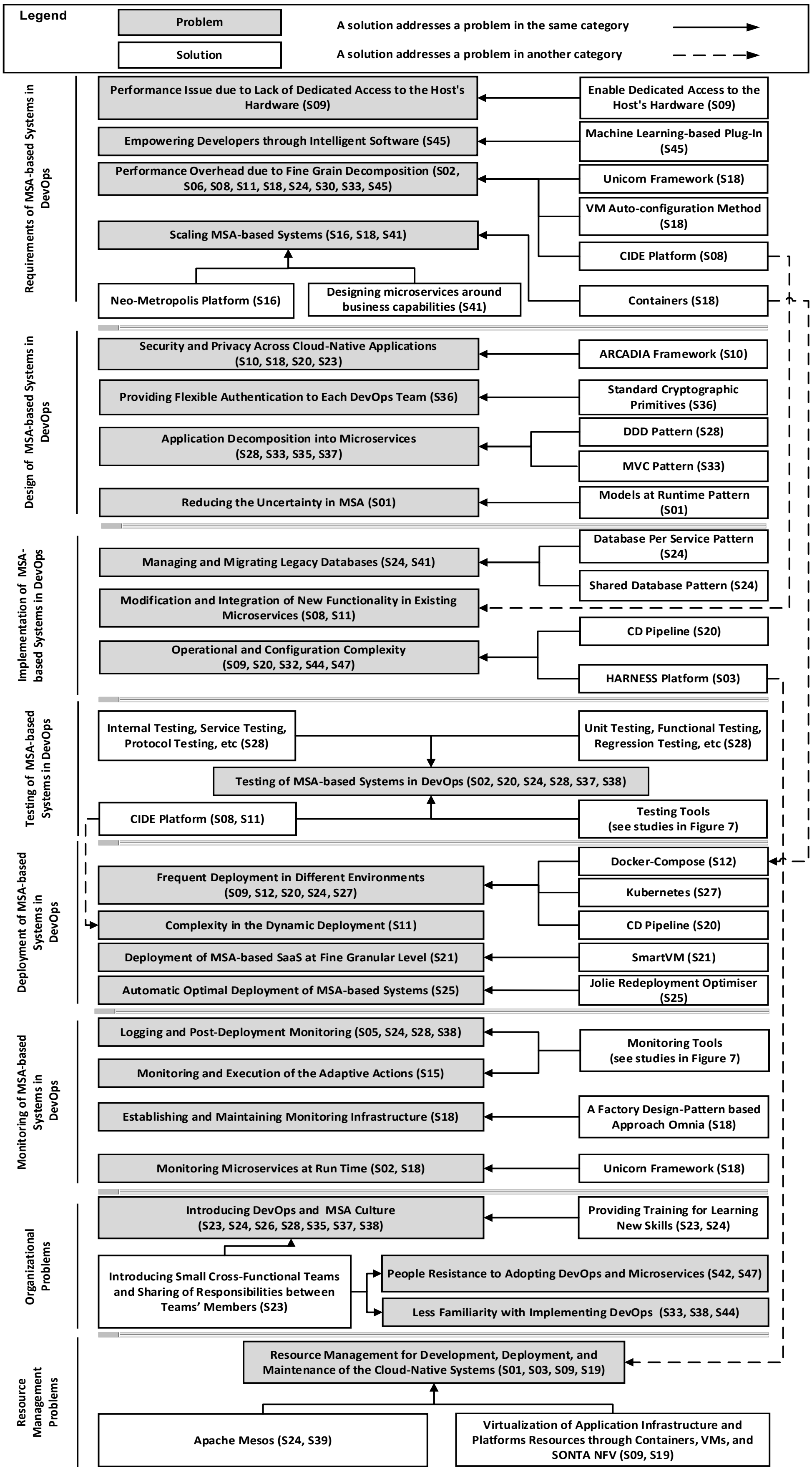

Figure 6. An Overview of Classification of Problems and Their Solutions from the Selected Studies

### 4.2.2. RQ2.3: Research Challenges

This section presents the identified challenges from the selected studies to answer RQ2.3 "*What challenges have been reported while implementing MSA in DevOps?*". It should be noted that we consider the problems that do not have any solution in the selected studies as "research challenges" (see Table 8). We briefly present these challenges below.

**Ch1: Performance issues due to frequent communication**: This challenge has been reported in three studies (S24, S30, S45). We identified different reasons for performance issues. For instance, Study (24) reports that communication between too fine-grain microservices introduces the complexity, with which the performance of the MSA based systems decreases. Study (S30) highlights that inter-process communication between microservices through synchronous HTTP could be the reason for performance issues, and Study (S45) indicates that performance issues might also occur when microservices interactions are routed through third-party solutions.

**Ch2: Providing security at runtime**: Threats related to the security of MSA based systems increase due to virtualization of microservices through containers and VMs. Therefore, special care about privacy and security is required at design and runtime (S35). Also, Study (S36) reports that boundaries of MSA based systems are protected through authentication methods (e.g., passwords) and transport layer security. Generally, the security of individual microservices tends to be neglected, leading to security vulnerabilities at runtime for MSA based systems (S36).

**Ch3: Generating runtime architectural models**: Requirements for dynamic systems (e.g., MSA based systems) are often necessary to be represented through runtime architectural models, which help in decision-making process for adaptive system development (e.g., MSA based systems). However, generating runtime architectural models from development models and abstract representation of these models through Model-Driven Engineering (MDE) methodology for continuous development is challenging. Study (S02) reports that performance modeling at runtime and deployment for MSA based systems is challenging due to shifted use cases, the lack of modelling abstractions, and keeping the models up-to-date automatically.

Table 8. Research challenges in the Context of MSA in DevOps

| Challenges | Key Points and Selected Studies |
|---|---|
| Ch1: Performance issues due to frequent communication | ▪ Performance overhead due to frequent inter-process communication of microservices (S24, S30, S45) |
| Ch2: Providing security at runtime | ▪ How to deal with privacy and security at design and runtime for MSA based systems (S35)<br>▪ How to secure individual microservice (S36) |
| Ch3: Generating runtime architectural models | ▪ How to introduce self-adaptive architecture for cloud-based applications (e.g., microservices) (S01)<br>▪ How to generate performance models at runtime for MSA based systems (S02) |

## 4.3. MSA Description Methods, Patterns, and Quality Attributes

This section presents the results of RQs about MSA description methods (RQ3.1), MSA patterns (RQ3.2), and QAs when employing MSA in DevOps (RQ3.3).

### 4.3.1. RQ3.1: MSA Description Methods in DevOps

We identified MSA description methods used for expressing, communicating, and analyzing MSA design in DevOps, to answer RQ3.1 "*What methods are used to describe MSA in DevOps?*". Table 9 shows 19 MSA description methods which are classified into five categories, including boxes and lines, UML, formal methods, Architecture Description Languages (ADLs), and Others.

The results show that most of the selected studies use informal boxes and lines for representing the high-level design, functional decomposition, and process flow of MSA based systems, and there are four description methods, architectural block diagram, functional flow block diagram, tiered architecture, and flowchart, in the boxes and lines category. The description methods under the boxes and lines category are used to describe MSA of various systems. For example, architectural block diagrams are used to describe the MSA of CIDE in Study (S08) and Study (S11), functional flow block diagrams are used to represent the scenarios of a DevOps based monitoring approach (i.e., OMNIA) for microservices in Study (S05), tiered architecture is used to illustrate MSA of High-Availability and Disaster Recovery (HADR) system in Study (S01), and flowcharts are used to show creation and execution of containerized microservices for IoT in Study (S14). We found that four types of UML diagrams are also used in the studies to represent different aspects of MSA based systems. For instance, activity diagrams (S13) are used to represent migration and development process flow of an MSA based mobile application (i.e., EasyLearn), sequence diagrams (S03) are used to describe object interactions of MSA based HARNESS platform, class diagrams (S46) are used to show the static structure of the tool for incremental integration of microservices, and component diagrams (S12) are used to reflect the physical structure of an MSA based system (i.e., Backtory) in DevOps.

Besides informal diagrams (e.g., boxes and lines) and semi-formal diagrams (e.g., UML), we identified four studies (S01, S29, S33, S45) that make use of formal methods to represent MSA. For example, Fuzzy logic model (S01) and MAPE-K loop (S01) are used to capture dynamic behavior of cloud-based systems (e.g., MSA based systems), π-Calculus is used to model cloud services (e.g., microservices) executed by VM clusters, and formal model architecture (S45) is used to describe the MSA of Continuous Development Intelligent Assistant (CDIA). Moreover, we found five ADLs used in three studies (S08, S11, S25) for describing MSA in the CIDE platform and JRO tool, including Caste-centric Agent-Oriented Programming Language and Environment (CAOPLE), Cloud Application Modeling and Execution Language (CAMLE), Specification Language for Agent-Based Systems (SLABS), Service Desiderata Language (SDA), and Jolie. CAOPLE (S08), CAMLE (S08), and SLABS (S08) are used in the CIDE platform for modeling, development, and testing of model-based microservices, whereas SDA (S25) and Jolie (S25) facilitate the JRO for automatic and optimized deployment of MSA based systems. Furthermore, Study (S13) used Entity Relationship Diagrams (ERD) to model the data view of the MSA based mobile application (i.e., EasyLearn) and Study (S33) employed Business Process Modeling Notations (BPMN) to conceptualize the VM auto-configurator architecture for MSA based systems.

Table 9. Classification of MSA Description Methods from the Selected Studies

| Category | Description Method | Study ID | # of Studies |
|---|---|---|---|
| Boxes and Lines | Architectural Block Diagram | S08, S09, S10, S11, S14, S18, S22, S24, S27, S31, S32, S33, S34, S36, S37, S40, S43, S44, S47 | 32 |
| | Functional Flow Block Diagram | S05, S06, S13, S15, S19, S21, S31 | |
| | Tiered Architecture | S01, S03, S15, S21, S28 | |
| | Flowchart | S14 | |
| UML | Activity Diagram | S13, S33 | 6 |
| | Sequence Diagram | S03 | |
| | Class Diagram | S46 | |
| | Component Diagram | S12, S27 | |
| Formal Method | Fuzzy logic model | S01 | 4 |
| | MAPE-K | S01 | |
| | Multi-layer Fuzzy Cognitive Maps | S29 | |
| | π -Calculus | S33 | |
| | Formal Model Architecture | S45 | |
| ADL | CAOPLE. Language | S08, S11 | 3 |
| | CAMLE. Language | S08 | |
| | SLABS Language | S08 | |
| | SDA | S25 | |
| | Jolie | S25 | |
| Others | ERD | S13 | 1 |
| | BPMN | S33 | |

### 4.3.2. RQ3.2: MSA Design Patterns in DevOps

To answer RQ3.2 "*What MSA design patterns are used in DevOps context?*", we identified 38 MSA design patterns from 19 studies. The set of identified MSA design patterns are presented in

Table 10. We found that a few studies discussed MSA design patterns. The most recurring design patterns when implementing MSA in DevOps are: Circuit Breaker (5 studies), Migration pattern (4 studies), followed by Observer (2 studies), Load Balancer (2 studies), Scalability (2 studies), and Deployment (2 studies). It is important to note that 30 MSA design patterns have been mentioned in only two studies (S12, S33). In Study (S12) and Study (S33), four different MSA design patterns are reported for service discovery, namely: Client-side discovery, Server-side discovery, Service registry, Self-registration, and third-party registration. These studies (S12 and S33) also reported MSA design patterns related to the decomposition of application into microservices (e.g. DDD), data management (e.g., Database per service), reliability (e.g., Circuit Breaker), and external API (e.g., API gateway). We found that a set of studies proposed new patterns and new approaches based on existing patterns. For instance, Study (S01) proposed patterns to deal with uncertainty in cloud-native architecture (e.g., MSA), namely: Quality Models at Runtime, Control-Based Feedback Loop, and HADR patterns, and Study (S05) presents a monitoring approach for MSA based systems based on Factory pattern.

Table 10. MSA Design Patterns used in DevOps Context from the Selected Studies

| MSA Design Pattern | Study ID | MSA Design Pattern | Study ID |
|---|---|---|---|
| Circuit Breaker | S02, S12, S24, S27, S33 | Migration pattern | S12, S27, S29, S38 |
| Observer | S11, S33 | Load Balancer pattern | S12, S33 |
| Scalability pattern, | S35, S42 | Deployment pattern | S12, S31 |
| Database per Service | S24, S33 | Strangler pattern | S47 |
| Factory | S05 | TOSCA pattern | S22 |
| DDD pattern | S12 | Three-Tier Architecture pattern | S33 |
| Server-Side Discovery | | Model View Controller pattern | |
| Client-Side Discovery | | API Gateway pattern | |
| Internal Load Balancer | | Reactor pattern | |
| External Load Balancer | | Inter-Process Communication | |
| Configuration Server | | Event Driven pattern | |
| Edge Server | | Message Broker pattern | |
| Containerized the Services | | Command Query Responsibility Segregation (CQRS) | |
| Performance pattern | | Service Discovery pattern | |
| Monitoring pattern | | Self-Registration pattern | |
| Tolerant Reader pattern | | Third-Party Registration | |
| Quality Models at Runtime | S01 | Resource Utilization pattern | S21 |
| Control-based feedback loop | | API Access pattern | |
| HADR pattern | | Clustering of Workload pattern | |
| Shared Database | S24 | Heat pattern | S41 |
| Layered Architecture | | Virtual system patterns | |

### 4.3.3. RQ3.3: Quality Attributes

To answer RQ3.3 "*What quality attributes are affected when employing MSA in DevOps?*", we first confirmed the presence of QAs in the selected studies, and then we investigated whether QAs are positively or negatively affected when employing MSA in DevOps. Various QAs are considered for software systems (e.g., as specified in ISO/IEC 25010 [34]), but not all QAs are equally important for MSA based systems. Thus, we decided to use those QAs that are discussed either in MSA or SOA context, e.g., [35, 36]. We selected 15 QAs that are popular in MSA and SOA context and identified the positive and negative effect on these QAs when using MSA in DevOps. The results show that most of the QAs are positively affected during the implementation of MSA in DevOps (see Table 11).

It should be noted that several studies reported more than one QAs and some of the studies discussed both the positive and negative effects of using MSA in DevOps on the QAs; for example, Study (S08) and Study (S14). The leading positively affected QAs are Deployability and Scalability, which have been mentioned in 41 and 32 studies respectively. Some other leading positively affected QAs are Performance, Maintainability, Monitoring, and Testability. These results indicate that MSA in DevOps brings significant benefits, including, independent scalability, flexibility to consume new frameworks, improved product quality, and zero downtime deployment.

We also investigated the QAs that are negatively affected when employing MSA in DevOps. We found that Security is the most negatively affected QA which is mentioned in 11 studies, suggesting that MSA may introduce more vulnerabilities than monolithic applications. This is because, for example, microservices run via HTTP and use vulnerable third-party components, which may expose them for hackers' attack.

Table 11. Quality Attributes Affected when Employing MSA in DevOps

| Effect | Description and Examples | Quality Attributes with Study ID |
|---|---|---|
| Positive | The combination of MSA and DevOps offers many benefits, e.g., a positive impact on QAs, and helps to achieve QAs requirements. Two examples of the positive impact on QAs when employing MSA in DevOps are given below:<br>(1) The deployability of microservices in DevOps can be managed dynamically (S01, S03).<br>(2) The combination of MSA and DevOps provides a high degree of scalability, availability, flexibility, and portability (S02). | **Deployability** (S01, S03, S05, S06, S07, S08, S09, S10, S11, S12, S13, S14, S15, S16, S17, S18, S20, S21, S22, S23, S24, S25, S27, S28, S29, S30, S31, S32, S33, S34, S35, S36, S37, S38, S40, S41, S42, S43, S44, S46, S47) |
| | | **Scalability** (S01, S02, S06, S07, S08, S09, S11, S12, S13, S15, S16, S17, S18, S19, S21, S22, S24, S25, S27, S28, S29, S31, S33, S34, S35, S36, S39, S40, S42, S44, S46, S47) |
| | | **Performance** (S01, S03, S04, S07, S08, S09, S11, S12, S15, S16, S18, S19, S21, S22, S24, S25, S28, S31, S33, S35, S36, S40, S42, S43, S44, S47) |
| | | **Maintainability** (S01, S03, S05, S06, S07, S09, S11, S12, S13, S15, S16, S19, S21, S22, S29, S30, S31, S32, S33, S35, S36, S37, S41, S42, S43, S44, S47) |
| | | **Monitoring** (S01, S02, S04, S05, S07, S08, S09, S10, S11, S12, S15, S16, S17, S19, S21, S22, S25, S31, S32, S37, S40, S41, S42) |
| | | **Testability** (S02, S03, S04, S07, S08, S11, S12, S13, S14, S16, S17, S20, S22, S23, S28, S32, S33, S34, S35, S42, S44, S47) |
| | | **Flexibility** (S01, S07, S08, S09, S11, S13, S14, S15, S18, S19, S22, S23, S27, S28, S33, S35, S36, S40, S41, S43) |
| | | **Availability** (S01, S02, S04, S07, S09, S16, S20, S22, S26, S27, S28, S31, S32, S35, S37, S41, S42, S43, S46) |
| | | **Efficiency** (S01, S03, S07, S08, S09, S11, S15, S16, S21, S22, S27, S31, S33, S35, S38, S41, S42, S43, S44) |
| | | **Security** (S04, S07, S10, S16, S19, S22, S23, S33, S35, S36) |
| | | **Portability** (S02, S12, S13, S18, S22, S27, S40) |
| | | **Reliability** (S07, S11, S19, S29, S35, S47) |
| | | **Compatibility** (S20, S32, S40, S41, S46) |
| | | **Modifiability** (S07, S10, S20, S28, S46) |
| | | **Usability** (S35) |
| **Negative** | The combination of MSA and DevOps also brings certain issues, e.g., negative impact on QAs. Two examples of the negative impact on QAs when employing MSA in DevOps are given below:<br>(1) Accessing host resources for (micro)services that run over the containers can jeopardize the security of the host system (S09).<br>(2) Degradation of network performance when the number of containers is increased (S06) | **Security** (S09, S14, S21, S22, S23, S33, S35, S36, S43, S45, S47) |
| | | **Performance** (S02, S06, S08, S18, S22, S24, S30, S35, S43) |
| | | **Scalability** (S18, S21) |
| | | **Reliability** (S37, S44) |
| | | **Availability** (S21) |
| | | **Compatibility** (S37) |
| | | **Maintainability** (S33) |
| | | **Modifiability** (S03) |
| | | **Usability** (S05) |

## 4.4. Tool Support and Application Domains

This section reports the tool support for MSA based systems in DevOps (RQ4.1) and application domains that exploit the combination of MSA and DevOps (RQ4.2). We identified 50 tools and 11 domains from the selected studies. Details of the identified tools and application domains are presented in the subsections below.

### 4.4.1. RQ4.1: Tool Support

To answer RQ4.1 "*What tools are available to support MSA in DevOps context?*", we first collected the functionalities of the tools that are proposed or used in the selected studies. Then we classified the tools into seven categories (see Figure 7) based on the collected features.

We identified a wide variety of tools that support the development of MSA based systems in DevOps. The most popular tooling category is "Security Services and Tools" in which 14 security tools and services are reported. The second prominent category is "Monitoring Tools", which includes 11 tools. The "Version Control Tools" category has the least number of tools (two tools). We observed that GitHub as a version control system and Jenkins as an integration server are the most popular tools used in the selected studies.

The seven categories of tools are explained below:

- **Security Services and Tools**: Microservices provide public interfaces, use network-exposed APIs for communicating with other services, and are developed by using polyglot technologies and toolsets that may be un-secure (e.g., DevOps tools that do not fulfil security requirements). These make microservices to be a potential target for cyber-attacks; therefore, the security of MSA based systems demands serious attention. Without considering security, a combination of cloud, DevOps, and containerized MSA based systems may increase security risk for organizations striving to use these technologies [37]. We identified eight studies in which 14 security services and tools are reported. These services and tools provide different levels of security. For example, SONATA (S10, S19, S43) protects the development lifecycle of MSA based systems.
- **Monitoring Tools**: The highly dynamic nature of MSA based systems requires a robust monitoring infrastructure to diagnose and report faults and performance issues. We identified 11 tools from eight studies that can help to monitor MSA based systems in DevOps. JIRA is a general-purpose issue and bug tracking tool that can also be used to monitor MSA based systems. For example, JIRA helps to create a project for each microservice that provides enough autonomy for DevOps teams to control issues and bugs related to each microservice.
- **Continuous Integration Tools**: Normally, the code of MSA based systems needs to be frequently integrated into a shared repository. Under this category, we gathered 7 tools (see Figure 7) that are used to automate the continuous integration for MSA based systems. For example, Jenkins is an open-source automation server that is used to integrate various operations (e.g., build, test, and deployment) related to development of MSA based systems. Jenkins can also be used to automate DevOps practices (e.g., CI, CD).

- **Testing Tools**: Testing MSA based systems is a challenging task due to multiple independently deployable microservices. We identified four studies (S13, S14, S42, S44) that employed 6 testing tools to provide support for microservices testing strategies (e.g., unit testing, integration testing, end to end testing). For instance, Junit (S13) can be used for both unit and integration testing of MSA based systems, and Cucumber-Selenium (S13) can be used for end to end testing of MSA based systems.
- **Configuration Management Tools**: Managing configurations of MSA based systems is a challenging task because many microservices are frequently deployed and update their configuration files. Under this category, we report the 5 configuration management tools that can address configuration issues of MSA based systems. We found that Puppet and CHEF are the two most reported tools in the studies. Both tools are easily scalable and can effectively manage the configuration of development, testing, and monitoring infrastructure for MSA based systems. Puppet uses Puppet Domain Specific Language (DSL) and CHEF uses Ruby DSL to centralize and automate configuration management of microservices [38].
- **Build Tools**: Build tools are used to automate the process of converting source code files into application binaries for a software release in operational environment. We identified 4 tools in this category that are used to automate the build process of MSA based systems in DevOps. For example, Study (S13) and Study (S37) used the Gradle tool to automate the build process of MSA based systems. It is recommended that a combination of Spring Boot and Gradle could be the right choice for developers to develop and build MSA based systems in DevOps [39].
- **Version Control Tools**: To keep track of changes and releases, developers use version control systems. Under this category, we identified two tools (i.e., GitHub and Bitbucket) from the studies that are used for managing versions of MSA based systems. GitHub provides a convenient mechanism for implementing CI/CD pipelines for MSA based systems [40]. It also provides specialized languages (e.g., Ballerina) to support the development and deployment of microservices. Another tool is Bitbucket that provides not only version control, but also project planning, collaboration on code, continuous integration, continuous testing, and continuous deployment for MSA based systems in DevOps

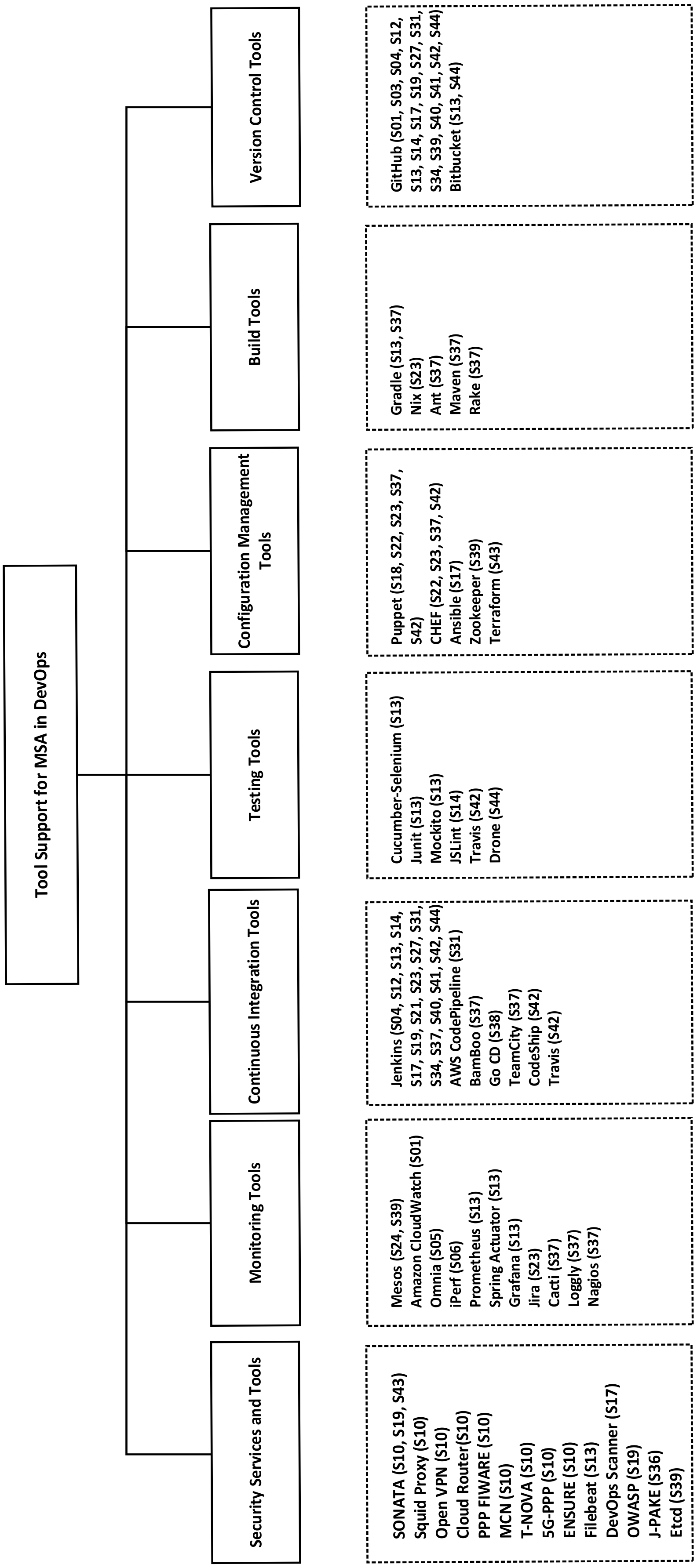

Figure 7. An Overview of the Tools for MSA based Systems in DevOps

### 4.4.2. RQ4.2: Application Domains

This section presents the results of the identified application domains from the selected studies. To answer the RQ4.2 "*What are the application domains that exploit the combination of MSA and DevOps*?", we used the data item (D14) from Table 5. We have characterized 9 the application domains after analyzing the functionality of systems that were proposed or investigated in the selected studies (see Table 12) in order to exploit the combination of MSA and DevOps. We observed that 15 (31.91%) studies did not provide any specific information regarding application domains. Therefore, we categorized those studies as "Not Mentioned".

Our results show that the "Software Development Tools and Framework" application domain has gained the most attention for implementation of MSA in DevOps, followed by "Telecommunication" and "Mobile Software".

Table 12. Application Domains that Exploit the Combination of MSA and DevOps

| Application Domains | Study ID |
|---|---|
| Not Mentioned | S02, S22, S24, S26, S28, S29, S31, S35, S37, S38, S41, S44, S45, S46, S47 |
| Software Development Tools and Framework | Microsoft HADR (S01), OpenStack (S01, S09), HARNESS (S03), Omnia (S05), CIDE (S08, S11), Unicorn Framework (S18), Azure PowerShell (S23), Azure Visual Studio Team Services (S23), Mobile SDK (S27) |
| Telecommunication | Cisco's Intercloud Analytics platform (S07, S16), Mobile 5G networks (S32, S43), SONATA NFV (S19), SONATA SDK (S43) |
| Mobile Software | Backtory (S12), Easy Learn (S13), Mobile SDK (S27), Mobile WebShop (S36) |
| E-Commerce System | Electronic Commerce (S17), Retail Application (S21), Mobile WebShop (S36) |
| Embedded System | Edge Devices (S06), SmartX IoT (S14), Smart Energy IoT (S34) |
| Financial Software | CRM (S33, S42), Finance (S42) |
| Healthcare Software | Remote Patient Monitoring (S10) |
| Webserver | Server Side as a Service (S27) |
| Distributed System | Connected Car (S39) |
| Others | Autonomic Management System (S15), Betting and Gaming (S20), Web Blog (S25), eServices Developments (S30), Container Management System (S40), Content Management (S42), Software for non-profit (S42) |

# 5. Discussion

In the following, we analyze and discuss the key findings of our study, along with their implications for research and practice.

## 5.1. Analysis of the Results

We further analyze and synthesize the results of our RQs related to research classification, problems, solutions, challenges, MSA description methods, MSA patterns, QAs, tool support, and application domains for the implementation of MSA in DevOps.

### 5.1.1. Research Status and Themes

We limited our search to the peer-reviewed literature from January 2009 to July 2018. The year 2009 was chosen as the terms MSA and DevOps were coined around the year 2009. We found an upward trend in the number of studies on MSA in DevOps context (see Figure 3), indicating researchers and practitioners are paying more attention to MSA in DevOps context. We noticed that 43 papers (91.48%) were published from January 2016 to July 2018. Before the year 2015, we did not find any study that discusses problems, solutions, challenges, description methods, patterns, QAs, tools, and application domains of employing MSA in DevOps context. Our findings show that conference & symposium papers are the most popular publication type with 23 studies (i.e., 48.29%), followed by journal articles (12 studies, 25.53%) (see Figure 4). One potential reason is that MSA and DevOps are fast-changing areas, and the work submitted to conferences and symposiums can get quick feedback and publication.

The 47 studies were published in 41 venues, and most of the publication venues (34 out of 41) of the studies belong to the "Internet, Cloud, and Services Computing" and "Software Engineering" categories (see Table 6). This result indicates that the combination of MSA and DevOps has been a broad research topic across various publication venues and computing disciplines (e.g., cloud computing, software engineering, telecommunication).

A systematic identification, naming, and classification of the research themes of MSA in DevOps is provided in Figure 5 and Table 7. As shown in Figure 5, the most recurring subthemes are Tools (13 studies, 27.65%), Approaches (12 studies, 25.53%), and Development and Deployment (12 studies 25.53), followed by Design (10 studies, 21.27%) and Testing (7 studies, 14.89%). These results indicate that researchers are not only putting efforts for proposing approaches and tools (e.g., CIDE, SMART VM, OMINA) to support MSA based systems in DevOps, but also examining the development lifecycle of MSA in DevOps context (e.g., design, implementation, testing, and deployment). Regarding requirements engineering, we only find some QAs in the selected studies (see Table 7). We did not find any research effort about requirements engineering practices or activities (e.g., requirement analysis, modeling, evaluation, etc.). This indicates that there is a lack of research from a requirement engineering perspective regarding MSA in DevOps, which is important as the input of architecting. Regarding "Development and Deployment" subtheme, we observed that combination of MSA and DevOps is successfully used for developing and deploying cloud-native systems (S01, S41, S42), enterprise systems (S17, S33), eServices (S30), and IoT-based systems (S14), indicating this combination (i.e. MSA and DevOps) can be used to build small to large-scale enterprise systems (see Table 7). We also observed that the research around the design aspects of MSA in DevOps is only limited to proposing reference MSA (S09, S22, S44), MSA patterns (S01, S13, S28), and MSA tactics (S07) to some extent. There is no research on other design topics (e.g., MSA analysis, MSA synthesis, MSA evaluation). We found 7 studies (S02, S13, S28, S31, S33, S34, S41) about testing MSA based systems in DevOps. The research effort in these studies is limited to brief introduction of testing strategies (e.g., S13, S28, S34, S41), testing experiences (e.g., S31, S33, S41), and performance testing challenge (e.g., S02) for MSA based systems. Moreover, monitoring MSA based systems is another research theme that is not explored adequately. We identified only 4 studies that present monitoring approaches (e.g., S05), frameworks (e.g., S15, S18), and performance monitoring challenges (e.g., S02). Therefore, more research effort on testing and monitoring MSA based systems in DevOps is required. Furthermore, regarding migration to MSA in the context of DevOps, we found only 7 studies, in which 4 studies (S12, S13, S27, S29) report the experiences of migration from monolithic systems to MSA based systems and 3 studies (S24, S38, S47) discuss the motivation and challenges regarding the migration. This observation reveals that the experiences and lessons learned during the migration from monolith to MSA when adopting DevOps are rarely explored.

### 5.1.2. Problems and Solutions

A one-to-one mapping between problems and solutions is provided in Figure 6. The identified solutions consist of MSA design patterns, guidelines, frameworks, development platforms, tools, etc. We observed that only a few problems were reported regarding QAs (i.e., performance, scalability, security) of MSA based systems. We did not find problems and solutions related to other QAs (e.g., availability, reusability, reliability, maintainability, modularity, portability), for example, problems and solutions related to improving reusability of existing microservices for new microservices in MSA based systems.

The proposed solutions for the design problems of MSA based systems in DevOps are patterns and frameworks. For instance, DDD and MVC patterns are recommended for decomposing an application into microservices, Model at Runtime pattern is proposed for reducing uncertainty in cloud-native architecture (e.g., MSA), and the ARCADIA framework is used to address security issues in MSA based systems. Our observation shows that (1) problems and solutions regarding the design of MSA based systems are only limited to few aspects (e.g., application decomposition, reducing uncertainty), (2) the role of DevOps in MSA design has not been explored thoroughly, and (3) MSA architecting activities (e.g., analysis, synthesis, evaluation, implementation, maintenance, evolution) in DevOps are not investigated. Regarding implementation (e.g., coding) of MSA based systems in DevOps, we found only a few problems reported that are mainly related to integrating microservices and migrating relational databases of monolithic systems to MSA based systems. We did not find any study that especially explored problems related to coding of MSA based systems in DevOps, and we believe that investigating problems and solutions from the coding perspective could provide a significant contribution to the literature related to the implementation of MSA based systems in DevOps.

Several studies (S02, S20, S24, S28, S37, S38) reported that testing of MSA based systems in DevOps is difficult, but we did not find any detailed discussion on how and why testing of MSA based systems is difficult in DevOps? Moreover, recommended strategies to test (e.g., unit testing, functional testing, regression testing) MSA based systems in DevOps are the same to testing monolithic systems. We did not find any testing strategy that is specifically designed to test MSA based systems in DevOps. The selected studies also indicate that there is no study conducted in the context of testing MSA based systems in DevOps.

DevOps infrastructure is expected to ease the deployment of MSA based systems, and our results show that there are several solutions (e.g., Docker-compose, Kubernetes, Smart VM, JRO) available to address the deployment problems of MSA based systems in DevOps. These solutions mainly address the problem of frequent and optimal deployment of microservices in DevOps. Some of these solutions are popularly and frequently used in the industry, for example, Docker-compose and Kubernetes.

Monitoring of MSA based systems is also considered as problem by seven studies (see Figure 6). Several studies reported logging and post deployment (S05, S15, S24, S28, S38) and monitoring and execution of the adaptive actions (S15) as problems without recommending any solutions. We investigated the features of the monitoring tools (see Figure 7) and some of them (e.g., Amazon CloudWatch, Jira, Loggly) are recommended as solutions to monitoring problems, for instance, Amazon CloudWatch can address the logging, post-deployment monitoring, and adaptive actions monitoring by collecting operational data in the form of logs, metrics, and events, and can also discover system-wide performance issues and take automated actions (e.g., troubleshoot) to keep running MSA based systems smoothly.

On the other hand, we observed that the proposed solutions in terms of frameworks, platforms, and tools are multifaceted, meaning that these solutions not only address the problems identified in the studies (see Figure 6), but also provide services for other software engineering activities (e.g., development, testing, deployment, and monitoring), for example, DevOps-based ARCADIA framework supports efficient and flexible development and operation of MSA based cloud applications (S10), and also allows developers to select multi-vendor solutions to secure the microservices development lifecycle. Unicorn framework is not only used to address performance, monitoring, and security issues, but also used to develop MSA based systems (S18).

### 5.1.3. Challenges

We initially identified several challenges that are presented without solutions in the selected studies, but after reviewing other selected studies, we found the solutions for most of the challenges, and consequently treated them as problems. But we could not find any solutions for three challenges (see Table 8). Possible reasons in this regard are that, technology is not mature enough to address those challenges, solutions do not exist, or authors did not mention solutions explicitly. We briefly discuss the identified challenges below.

Performance issue due to frequent communication between microservices is one of the challenges, and this challenge occurs due to several reasons, such as, asynchronous requests, third-party requests, selection of inappropriate databases (e.g., shared database over database per service), spreading service requests across multiple databases, poorly established database connection pool, etc. None of the selected studies proposes dedicated techniques and tools to address those reasons behind performance issues due to frequent communication. However, a few studies (e.g., [41-45]) discuss the performance of the Docker containers and VMs for the implementation of microservices. Salah et al. experimented and found that the performance of the services that are deployed on Amazon EC2 container is worse than services that are deployed on Amazon EC2 VMs [41]. It is mainly due to the reason that Amazon EC2 container does not run directly on physical hosts. Kratzke's experiment findings demonstrate that network performance degradation due to containers is not negligible, and can be improved by 10 to 20 percent by providing VM-based Software Defined Virtual Networks (SDVN) router applications directly on the host [42]. Potdar et al. assessed the performance of Docker containers and VMs, and found that Docker containers performed better than VMs [43]. Hence their findings contradict the findings of [41, 42]. Amaral et al. show that nested containers have no significant performance impact compared to regular containers [44]. The study by Amaral et al. also claims that the containers can be booted much quicker than the VMs. Goethals et al. found that Unikernels outperform Docker containers in many ways [45]. For example, Unikernels consume less memory than Docker containers when the number of deployment instances is less. The results of the above-discussed work indicate that performance issues may not be addressed by merely selecting the Docker containers, VMs, or Unikernels. Many other factors also need to be considered while selecting the strategies for implementing MSA, such as the number of services per host, development language, and cloud platform (e.g., Amazon Web Services, Google Compute Engine, Microsoft Azure).

Security is another common concern during the implementation of MSA. We identified some studies that discuss solutions for microservices security (see Figure 6). However, we did not find any solution for addressing microservices security at runtime (see Table 8). As discussed in [46], there are many reasons why security is challenging for containerized microservices, for instance, container deployment speed, small scale microservices that make complex access control rules and increased data traffic, and cloud-based environments. It is claimed that there are not many mature solutions available to address security issues of MSA based systems [46].

Final challenge that we observed is about generating runtime architectural models for MSA based systems. Runtime architectural models can help in making runtime decisions about dynamic changes for executing MSA based systems (see the reasons in point 3 of Section 4.2.2). Runtime architectural models can also help to understand microservices behavior during execution, and contribute to runtime responses to changes, recovery, and evolution of MSA based systems in DevOps.

### 5.1.4. MSA Description Methods

Modeling diagrams (boxes and lines, and UML), formal methods, ADLs, and other description methods as listed in Table 9 are used to describe different aspects of MSA based systems. The results show that Architectural Block and Functional Flow Block diagrams were extensively employed to describe the high-level view of the MSA. The results reported regarding the use of UML diagrams (e.g., class, sequence, component diagrams) indicate that logical, process, and implementation views of MSA based systems have been represented using the UML diagrams while deployment views are not. However, we did not find any clear justification in the studies about the choice of using these UML diagrams in describing MSA views. The use of ADLs (e.g., CAOPLE) and formal methods (i.e., Fuzzy model) suggests that MSA is not only described by graphical models but also by formal languages and notations. For instance, CAOPLE is an agent-oriented programming language that can be used to model MSA in CIDE, which is a DevOps supported IDE for developing MSA based systems.

It is essential to mention that except for ADLs (e.g., CAOPLE), we did not observe any difference between the description of MSA and that of other types of architectures (e.g., monoliths, SOA). For instance, architectural block diagrams, tiered architecture diagrams, and class diagrams can be used to describe the architecture of monolithic applications, SOA, and MSA based systems. We found that the selected studies did not mention the reasons behind selecting certain description methods for MSA in DevOps. In line with Di Francesco et.al. [47], we argue that this concern can be partially addressed by working on a standard ADL for describing MSA.

### 5.1.5. MSA Design Patterns

This SMS reports 38 MSA design patterns in DevOps context, which are identified from 19 studies (40.42%) as listed in Table *10*. We observed that Circuit Breaker (5 studies, 10.63%) is the dominant MSA pattern, indicating that the cascading failure is a major concern when implementing MSA in DevOps. Another prominent MSA pattern is the Migration pattern (4 studies, 8.51%) that recommends the best practices (e.g., enabling CI, recovering the existing architecture) for the migration from monoliths to MSA in DevOps. We realized that several MSA patterns could be associated with the software development activities of MSA based systems, such as DDD (S12) and MVC (S33). Deployment patterns suggesting that microservices can be deployed by using Service Instance Per Container (S12), Multiple Service Instances Per Host (S31), and Service Instance Per Host (S31) patterns. Patterns for databases suggest that a database for microservice can be manage by using Database Per Service (S24, S33) and Shared Database (S24). The application of Database Per Service pattern is considered as an effective way to achieve loosely coupled-MSA based systems [48], and the application of Shared Database pattern is suitable for the situation where multiple microservices need to access persist data owned by other services [49]. The SMS results reveal several other patterns that are recommended for communication, monitoring, and service discovery in MSA. The examples of such patterns are API Gateway, Factory Design, Service Discovery, and Inter-Process Communication. The analysis of identified patterns shows that many MSA patterns are borrowed from monoliths and SOA. Examples of such patterns are MVC, Observer, and Load Balancer, and these patterns can also be effectively employed in the design of MSA. For instance, the Observer pattern can be used to observe the state changes of resources via RESTful API, and the Load Balancer pattern can accommodate the increasing load on services.

However, it is worthwhile to mention that most of the identified MSA patterns exclusively address the issues in microservices context, and except for the Migration patterns presented in Study (S12), we did not find any other patterns specifically proposed for MSA in DevOps. To this end, further research effort is required to propose new patterns to facilitate MSA in DevOps, for example, patterns to support the microservice CI/CD pipeline.

### 5.1.6. Quality Attributes

Table 11 shows that Deployability (41 studies, 87.23%), Scalability (32 studies, 68.08%), Performance (27 studies, 57.44%), and Maintainability (27 studies, 57.44%) are the most frequently mentioned QAs, while other QAs (i.e., Availability, Compatibility, Codifiability, Efficiency, Flexibility, Monitoring, Portability, Security, Reliability, Testability, Usability) are also frequently reported. One potential reason these QAs are extensively reported is that MSA and DevOps have a significant influence on QAs. To understand this influence, we further analyzed the positive and negative effect of the MSA in DevOps context on QAs. The results show that almost all the mentioned QAs are positively affected, suggesting that the combination of MSA and DevOps offers a significant improvement in QAs. For example, MSA can help to achieve scalability and performance along with increased efficiency and availability of MSA based systems. On the other hand, DevOps enables software companies to establish a safe and reliable infrastructure to improve the testability, monitoring, and deployability of microservices.

Several studies also report the negative effect of the combination of MSA and DevOps on QAs (e.g., Security, Performance, Scalability, Reliability), suggesting that employing MSA in DevOps may also bring some drawbacks. For example, microservices pose security challenges due to inter-service communication over the distributed network. It is worth mentioning that all the negatively affected QAs are also reported as the positively affected QAs in several selected studies (see Table 11), meaning that certain QAs can be affected both positively and negatively. For example, Study (S08) reported that MSA provides better performance than monoliths, whereas Study (S24) mentioned that MSA can introduce a performance overhead if the communication between microservices is done using RESTful APIs over HTTP. We also observed that many aspects related to QAs are not explicitly presented in the selected studies. For example, how to specify QA requirements for MSA in DevOps, how to achieve QAs through the tactics proposed for MSA in DevOps, and how to evaluate the QAs of MSA in DevOps.

### 5.1.7. Tool Support

Regarding tool support for MSA in DevOps, we identified 50 tools and classified them into 7 categories (see Figure 7). We found that "Security Services and Tools" (14 tools) and "Monitoring Tools" (11 tools) are dominating categories, meaning that a wide range of tools and services are available to protect and monitor MSA based systems in DevOps. We identified several security tools (e.g., Sandbox, Open WRT, DevOps Scanner) that can be used to secure microservices. However, we made two major observations regarding identified security tools: (1) there is a lack of detail about how these tools can be used to secure microservices in DevOps and (2) security has not been sufficiently explored for microservices in DevOps. Our SMS found some tools (e.g., Amazon CloudWatch, Omnia) that can be used to monitor microservices, but we did not find any discussion in the studies about what kind of monitoring (e.g., application monitoring, network monitoring, user-centric monitoring) these tools can perform, and how to use tools to automate the monitoring process for MSA based systems in DevOps. The continuous integration category highlights Jenkins as a popular tool for CI and CD pipeline of microservices. The results related to tool support indicate that there is a rich set of tools that provide support to various stages of MSA based system development in DevOps. For instance, version control of microservices (e.g., Bitbucket), build automation (e.g., Gradle), testing (e.g., Junit), configuration management (e.g., Chef).

Even though dozens of tools are available to support MSA in DevOps, we observed that automated decision support (e.g., AI-based) for security, monitoring, testing, configuration management, and build is rarely available in the identified tools. The analysis of the identified tools also reveals that except a few tools (e.g., CIDE, Omina), the industry contribution for tool support is a way ahead of the academic contribution. One reason could be that MSA and DevOps are originally proposed and practiced in industry, and they have an urgent need to promote CI, CD, monitoring, testing, and deployment of MSA based systems with tool support, which is critical for the maturity of new technologies in industry.

### 5.1.8. Application Domains

We observed that 15 studies (31.91%) did not provide any specific information regarding application domains, and the domains identified from the remaining studies can be classified into 9 categories. The dominating application domain is "Software Development Tools and Framework" suggesting that the combination of MSA and DevOps has been applied for two major purposes. First, it is used to develop the new IDEs (e.g., CIDE) and multi-cloud services framework (e.g., Unicorn framework). Second, it is used to evaluate the existing technology (e.g., Azure Visual Studio Team Services) choices for building microservices in DevOps. The results show that on the one side the combination of MSA and DevOps has been applied in emerging domains like Mobile 5G networks (S32, S43), Edge devices (S06), SmartX IoT (S14), Connected car (S39), and Mobile software; on the other side, a significant number of studies focus on traditional application domains, for example, CRM (S33, S42), Electronic commerce (S17), Retail applications (S21), Mobile Webshop (S36), and Betting and Gaming (S20). These results indicate that the combination of MSA and DevOps is growing in various application domains.

Although the results indicate that the implementation of MSA in DevOps has been applied to a wide range of application domains, there are still many application domains in which MSA in DevOps is seldom employed, e.g., Embedded systems, Financial software, and Healthcare software.

## 5.2. Implications for Researchers

(1) MSA and DevOps both have been emerged and continuously evolved in the industry. However, we found that 55.3% of the selected studies are originated from academia, and only 21.27% of the studies have an author from the industry. Therefore, we encourage academic researchers to collaborate more with the industry to fill this gap.

(2) As discussed in Section 5.1.1, we did not find enough evidence related to requirements engineering, architecture/design, development, and testing of microservices in DevOps. One possible reason for lack of research in these areas is that the MSA and DevOps combination is relatively new, and most of the studies emphasize proposing approaches and providing tool support for MSA in DevOps without empirically evaluating them. Therefore, there is a clear need for conducting empirical research in the following areas:

- How are MSA requirements engineering activities (e.g., elicitation, documentation, and validation) practiced in DevOps?
- What are microservices architecting activities (e.g., analysis, implementation, evaluation, description, maintenance, and evolution, etc.) practiced in DevOps context?
- How to model runtime aspects of microservices in DevOps context?
- How to customize existing ADLs for describing MSA in DevOps.
- What are the best DevOps practices for MSA based systems?
- How does DevOps support testing MSA based systems?
- How to meet security in individual microservices without compromising the autonomy of the DevOps teams?

(3) We did not find solutions for three challenges (see Section 4.2.2). Given the increased importance of performance, security, and generating runtime architectural models of MSA based systems in DevOps, there is a clear need for further research to explore the identified challenges, for example, how MSA based systems should be designed and implemented in DevOps to meet performance expectations and mitigate security issues, how to generate runtime architectural models for making decisions about dynamic changes in MSA based systems.

(4) The use of informal architectural description methods (e.g., boxes and lines) can introduce the inconsistency issues between system design and implementation of the system [50]. Based on the evidence gained from the selected studies, we found that most of the studies describe MSA design by using boxes and lines. Therefore, we suggest that it will be valuable to identify and analyze the pros and cons of the informal description methods. Moreover, further research is also needed to explore why formal description methods (e.g., ADLs) are rarely used when applying MSA in DevOps context.

(5) We found several MSA patterns that can be applied to the different aspects (e.g., development, deployment) of microservices. However, we also found that there is a lack of patterns that are exclusively created for and can be applied to MSA in the context of DevOps. Therefore, we encourage researchers and practitioners to investigate architecture and design (e.g., creational, structural, behavioral) patterns in the context of MSA in DevOps.

(6) Recently, much attention has been paid to "observability" in the MSA community. Observability has three pillars, i.e., logging, monitoring, and distributed tracing [51, 52]. Logging means maintaining log data (e.g., error, failure, state transformation) generated by applications and infrastructure. Monitoring is an action in which data metrics (e.g., resource usage, system availability, threads) are collected, aggregated, and analysed to maintain the overall health of systems, and distributed tracing is a technique with which all the trace data is used to give meaningful insight for a request across several systems [51]. However, the selected studies (e.g., S02, S05, S15, S18) from our SMS only discuss the monitoring of MSA. We emphasize that further research is required to investigate logging and distributed tracing in the context of MSA and DevOps.

(7) We did not find any study that has explored MSA in DevOps in terms of refactoring, evolution, or maintenance. Therefore, further research is needed to explore the potential challenges, best practices, and tools in the above-mentioned activities.

(8) Whilst a dozen of tools has been identified in this SMS, we did not find any study that reports explicitly how these tools support MSA in DevOps. Therefore, more empirical studies are needed to understand and evaluate the tools supporting MSA in DevOps.

## 5.3. Implications for Practitioners

(1) The research on MSA in DevOps is still a relatively new and unexplored research topic. Practitioners are encouraged to frequently report their experiences when applying MSA in DevOps. This helps to reduce the gap between academic research and practices.

(2) This SMS reveals that monitoring of 100s of independent microservices and diagnosing of a failed service among them is a challenging task. The matter gets worse in DevOps context as the deployment of microservices happens at an extremely accelerated pace. To alleviate this issue, practitioners need to find smarter solutions that align with DevOps for detecting and reporting the failed microservices in real-time.
(3) A higher number of microservices in an application demand more security solutions for system protections. We found several studies that present security solutions for MSA based systems. However, practitioners still need to work on more advanced solutions to handle security for a variety of operating systems, languages, and frameworks when implementing MSA in DevOps.
(4) Several issues are reported in the literature related to microservices performance degradation. The typical reason for the performance degradation is that microservices consume an enormous amount of resources across the network and the burden (e.g., network latency) on the servers with microservices is greater than monolithic applications. Therefore, practitioners need to develop solutions for minimizing and mitigating performance degradation issues.

## 5.4. Analysis of the Results through a Word Cloud

We generated a word cloud (see Figure 8) by considering the titles, abstracts, keywords, and conclusions of the selected studies. Common words such as "*and*", "*of*", and "*the*" were automatically removed by Wordle[1]. We also removed several unnecessary words like "*used*", "*new*", and "*work*" from the generated word cloud. The dominant terms are "microservice" and "DevOps" that reflect the topic of this SMS (i.e., microservices architecture in DevOps). Other dominant terms like "development", "operation", "design", "monitoring", "testing", "approach", "tool", and "migration" partially reflect the classification of research themes on MSA in DevOps (see Figure 5). The terms such as "challenge", "issue" are aligned with RQ2.1 and RQ2.3, indicating the problems and challenges of MSA in DevOps. This word cloud also highlights several terms related to the identified solutions (see Figure 6). For instance, "container", "Docker", and "SONATA". We found only two QAs (i.e., "performance", "scalability") discussed in the selected studies in the word cloud. Overall, the word cloud mainly highlights the classifications of the research themes, problems and solutions, and challenges in the context of MSA in DevOps.

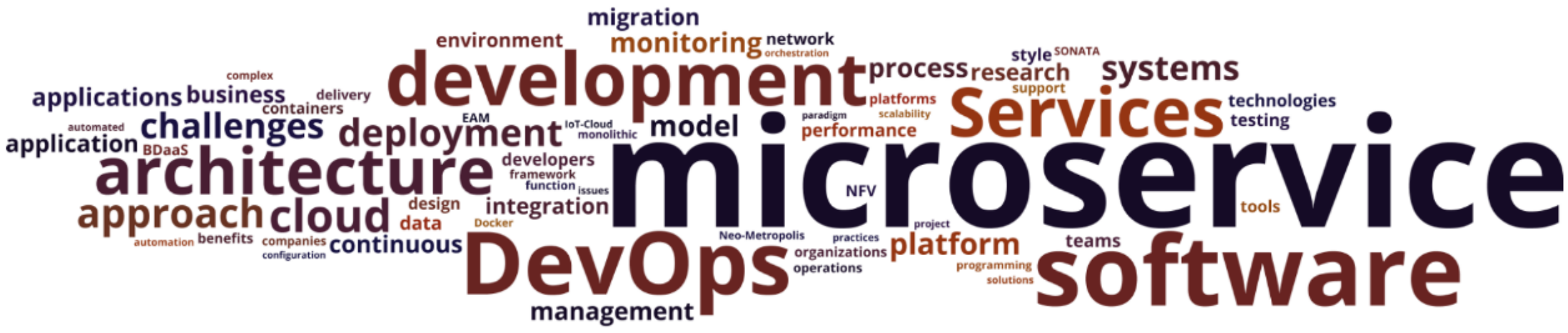

Figure 8. An overview in a word cloud of the key terms discussed in the selected studies

# 6. Threats to Validity

Several threats can affect the results of this SMS. To mitigate these threats, we followed the guidelines for conducting SMSs/SLRs [26, 27], and analyze the validity threats to our SMS according to four types of validity threats [53, 54]. In this section, we discuss the following validity threats associated with the different activities of this SMS.

## 6.1. Internal Validity

Internal validity refers to the factors that could affect the analysis of the data extracted from the selected studies. The threats to internal validity could happen in the following steps of this SMS:

[1] http://www.wordle.net/

**Study Search**: There is a possibility to miss relevant studies during the study search process. To mitigate this threat, we used the primary and snowballing search processes (see Section 3.2). To retrieve as many primary studies as possible through the primary search, we executed two search strings (i.e., String 1 and String 2) parallelly on seven popular databases. Additionally, we employed two other measures to minimize the threats in the search strategy: (i) The search strings were iteratively improved through pilot search before execution on the databases. During the pilot search, we found that many studies could not be retrieved through String 1. Therefore, we decided to use String 2. (ii) We also employed the backward and forward snowballing techniques to find the relevant studies in the last round of the search process.

**Study Selection**: We rigorously defined the study screening and selection process in Section 3.4. To avoid personal bias in study selection, we adopted a two-step process: (i) screening studies and (ii) qualitative assessment of the selected studies. During this process, we calculated the maturity of the study by using Formula 1 defined in Section 3.4.2. Furthermore, the first author of this study performed the screening of studies through explicitly defined criteria in Section 3.4.1 and Section 3.4.2, then the second and third authors of the study independently verified the screening results. It should be noted that all the researchers of this study have enough expertise, knowledge, and research experience about microservices and DevOps.

**Data Extraction**: Researchers' bias in data extraction can be a fundamental threat in any SMSs and SLRs. We mitigated this threat by creating the data extraction form (see Table 5) to extract data consistently. The data was initially extracted by the first author, and in case of any doubt about the extracted data, continuous discussions were organized between all the authors. As suggested in [54], a subset of the extracted data was verified by the second and third authors.

**Bias on Themes Classification:** Incorrect classification of the data and primary studies may bring subjective interpretation bias. To mitigate this bias, we employed the guidelines of thematic analysis technique proposed by Braun et al. [33].

**Data Synthesis**: We applied qualitative and quantitative methods to analyze the extracted data. The bias on synthesizing data may affect the interpretation of the results. To mitigate this threat, the synthesis of the collected data was performed through a well-established thematic analysis method for qualitative data and descriptive statistics for quantitative data.

## 6.2. External Validity

The threats related to external validity refer to the degree to which the results of a study can be generalized. The results of this SMS provide an overview of the state of research on MSA in DevOps context. Therefore, obtained results, analysis of the results, and conclusions drawn are only valid in the study topic. In order to achieve external validity, we developed the study protocol that rigorously specifies the whole process of conducting this SMS. In addition, to collect the relevant studies, we searched the peer-reviewed literature published between January 2009 and July 2018 in the seven most popular databases.

## 6.3. Construct Validity

Construct validity concerns whether correct operational measures have been taken for collecting the data to be studied. The primary constructs of this SMS are two concepts “MSA” and “DevOps”. The use of incorrect or incomplete search terms and inappropriate search strategies can bring threats like missing of relevant papers and inclusion of many irrelevant papers during the search process, and exclusion of relevant papers during the selection process. To alleviate these threats, we adopted the following operational measures: (i) We conducted the pilot search to ensure the appropriateness and completeness of the search terms. (ii) We used the seven most popular databases in computer sciences and software engineering research. We also customized each search string according to the peculiarities of the databases to obtain the relevant studies.

## 6.4. Conclusion Validity

Threats to conclusion validity are concerned with issues (e.g., inaccuracy of data) that affect the ability to reach the correct conclusions. We tried to mitigate these threats by applying the best practices (e.g., search protocol, pilot search, pilot data selection) proposed by Kitchenham et al. [26] and Petersen et al. [27]. In addition, to address the threats to conclusion validity, several brainstorming sessions were arranged among the authors about the interpretation of the results and conclusions.

# 7. Existing Systematic Reviews

We identified seven secondary studies [47, 55-59] that reported different aspects of MSA or DevOps. Five studies (including a systematic grey literature review) [47, 55-58] mainly focus on the MSA aspects, one study [59] present review on DevOps, and only one study [60] have presented the review on continuous architecting with MSA and DevOps.

Di Francesco et al. [47] reported the results of an SMS on MSA based on 103 primary studies published from 2008 to May 2017. This review presents the state of the art on microservices architecting with respect to the number of publications, research focus, and potential for industrial adoption. Major contributions of this SMS are (i) a framework for classifying, comparing, and evaluating the research on architecting microservices, (ii) evaluation of research results about MSA for industrial adoption, and (iii) a systematic map for current research and implications for future research. This SMS only focuses on architecting with microservices while confirms that there exists a close relationship between MSA and DevOps.

Felipe et al. [55] conducted a Systematic Literature Review (SLR) on MSA patterns and tactics. They have explored MSA patterns and tactics for microservices from 66 academics studies published from 2011 to 2017. They also investigated the association of the QAs with MSA patterns and tactics for microservices. Major reported findings are (i) the identification of 38 MSA patterns, (ii) confirmation about non-availability of tactics for MSA in academic literature, and (iii) recognition of scalability and performance as prominent QAs in MSA. This SLR claims MSA emerged from DevOps ideologies, and MSA requires to be explored from DevOps and Internet of Things (IoT) perspectives.

Alshuqayran et al. [56] reported an SMS on MSA based on 33 primary studies published from 2014 to 2016. The objectives of their research are to explore the existing architectural support for microservices and characterize a framework based on MSA challenges. This SMS reported MSA challenges, MSA diagrams/views, and the QAs related to MSA.

Taibi et al. [60] reported the results of an SMS on MSA based on 42 primary studies published from 2011 to 2016. The focus of this review is to characterize the MSA style patterns and principles in the context of DevOps. This study presents a list of MSA style patterns and the advantages, disadvantages, and lessons learned of each identified pattern. In addition, this study also reported the DevOps techniques classification scheme aligned with DevOps practices (e.g., planning, coding, and testing), for instance, semantic models for planning practice and agent-oriented programming language for coding practice.

Soldani et al. [57] conducted a Systematic Grey Literature Review (SGLR) on 51 resources (i.e., white papers, blog posts, and videos). The focus of this SGLR is to sort the gap between academic research and industrial practices on microservices. The main contribution of this study is the identification of technical and operational advantages and challenges, during the design, development, and operations of MSA based systems.

Pahl et al. [58] conducted an SMS on MSA based on 21 primary studies published in 2014 and 2015. This SMS aimed to identify, classify, and systematically compare the literature published on microservices and their application in the cloud. They organized and analyzed the selected studies into three types (i.e., methodological support, architecture support, and platform/tool support) by using a characterization framework. The results of this SMS show that (i) MSA style as an emerging trend with CD/DevOps culture and (ii) MSA based applications are mostly employed in cloud and containers.

Jabbari et al. [59] investigated 49 studies through an SMS to understand the DevOps definitions and practices, as well as to identify similarities and differences between DevOps and other software development methods. They explored the definitions of DevOps by considering "*central components*". The examples of the *central components* are development and operations, communication, collaboration, team working, bridging the gap, Continuous Integration (CI), and automatic deployment. To explore the DevOps practices, the authors identified the explicitly presented practices in literature and categorized these practices according to the software engineering knowledge areas (e.g., software engineering management, software construction, and software configuration management). The identified DevOps practices are continuous planning, feedback, monitoring, integration, deployment, and testing. This SMS also compared DevOps with agile, cloud computing, cloud management, waterfall, Information Technology Infrastructure Library (ITIL), and quality assurance.

Table 13. A Comparison of Selected Studies between this SMS and the Existing Secondary Studies

| Study | # Total Included Studies | Overlapped Studies | Type | Search Period |
|---|---|---|---|---|
| Di Francesco et al. [47] | 103 | 1 (S27) | SMS | From 2008 to 2017 |
| Osses et al. [55] | 66 | 1 (S02) | SLR | From 2011 to 2017 |
| Alshuqayran et al. [56] | 33 | 2 (S02, S09) | SMS | From 2014 to 2016 |
| Taibi et al. [60] | 42 | 1 (S27) | SMS | From 2011 to 2016 |
| Soldani et al. [57] | 51 | 0 | SGLR | From 2014 to 2017 |
| Pahl et al. [58] | 21 | 1 (S27) | SMS | From 2014 to 2015 |
| Jabbari et al. [59] | 49 | 0 | SMS | From 1969 to 2016 |

The existing secondary studies (see Table 13) and our work can be differentiated in the following aspects:

**1) Study Objective**

The SMS [47] presents the trends, focus of research, and potential for industrial adoption of architecting with microservices. The SMS [56] reported the challenges, diagrams/views, and QAs related to microservices systems. The SMS [60] characterized the MSA principles and patterns in DevOps context. The SMS [58] identified, classified, and compared the existing research on MSA and its application in the cloud. The SGLR [57] reported the industrial evidence about the technical and operational challenges and advantages related to the design, development, and operations of MSA based systems. The SMS [59] presents a review of DevOps definitions, practices, and similarity of the DevOps practices with other development methods. Except for SMS [60], all existing reviews (see Section 7) focus on either MSA or DevOps whereas our SMS intends to shed light on the combination of MSA and DevOps.

**2) Search Strings and Search Period**

Our search strings are different as we decided to combine the MSA terms with “DevOps” in the search strings. However, the search strings designed for other reviews either only covering MSA or DevOps. Moreover, previously reported reviews used only automatic search whereas we used both automatic search and snowballing. Out of the 47 selected studies, we identified two studies (S07 and S20) through the snowballing technique [30]. We searched the studies from January 2009 to July 2018, which makes our search is the most updated one.

**3) Research Questions**

Our six RQs are different from the existing published SMSs. For example, RQ1.1 and RQ1.2 aim to report the frequency of the research, type of the published literature, and classification and mapping of the research themes in an MSA and DevOps combination perspective. The RQs related to challenges (RQ2.1) and solutions (RQ2.2) for adopting MSA in DevOps context are unique. The RQs related to MSA description methods (RQ3.1), patterns (RQ3.2), and QAs (RQ3.3) are similar to some extent with RQs presented in other SMSs/SGLRs. For example RQ3.2 in our SMS and RQ1 and RQ2 (MSA patterns) in [55] respectively. Furthermore, RQ4.1 and RQ4.2 in this SMS have some overlaps with RQ3 (methods, techniques, and tool support to enable MSA development and operation) in [58]. However, our study explicitly explores the literature on MSA in DevOps to answer these RQs.

**4) Selected Studies**

We also compared the selected studies of our SMS with the selected studies of the existing secondary studies. We identified that only three selected studies (S02, S09, S27) of this SMS are included in other SMSs/SLRs/SGLRs (see Table 13, overlapped studies).

### 5) Study Results

Table 14 shows the comparison between the results of our SMS and the existing secondary studies (i.e., SMSs/SLRs/SGLRs). It is demonstrated in Table 14 that the results of our SMS are significantly different from the existing reviews. For instance, the existing reviews do not provide thematic classification, problems and their solutions, challenges, positive and negative impact on QAs, tools, and application domains in the context of MSA and DevOps combination. For example, Pahl and Jamshidi [58] identified the existing research issues about microservices, which are different from the challenges identified in our SMS in the context of MSA in DevOps. However, our findings related to MSA description methods and MSA design patterns have, to some extent, an overlap with the results of existing secondary studies. For example, some MSA description methods (e.g., UML diagrams) identified in this SMS can also be found in the SMS by Di Francesco et al. [47] and Alshuqayran et al. [56]. Similarly, some MSA design patterns (e.g., circuit breaker, API gateway, database per services, shared database) identified in our SMS can be seen in the review studies by Di Francesco et al. [47], Osses et al. [55], Taibi et al. [60], Soldani et al. [57], and Pahl and Jamshidi [58].

Table 14. A Comparison of Results between this SMS and the Existing Secondary Studies

| This SMS Results | Existing Secondary Studies | | | | | | |
|---|---|---|---|---|---|---|---|
| | [47] | [55] | [56] | [60] | [57] | [58] | [59] |
| Thematic classification (Section 4.1.4) | - | - | - | - | - | - | - |
| Problems and solutions (Section 4.2.1) | - | - | - | - | - | - | - |
| Challenges (Section 4.2.2) | - | - | - | - | - | - | - |
| MSA description methods (Section 4.3) | ✓ | - | ✓ | - | - | - | - |
| MSA design patterns (Section 4.3.2) | ✓ | ✓ | ✓ | ✓ | ✓ | ✓ | - |
| Quality attributes (Section 4.3.3) | - | - | - | - | - | - | - |
| Tools (Section 4.4.1) | - | - | - | - | - | - | - |
| Application domains (Section 4.4.2) | - | - | - | - | - | - | - |

# 8. Conclusions

This SMS provides the state of research about MSA in DevOps concerning research themes, challenges, solutions, MSA description methods, QAs, tools, and application domains. We selected 47 studies after a comprehensive search and selection process for data analysis. The key findings of this SMS can be summarized as follows:

(1) The increasing number of publications on MSA in DevOps context shows that this research topic is continuously gaining significant attention from the research community because 47 studies were published in 3 and a half years (From 2015 to July 2018).

(2) The reported research on the combination of MSA and DevOps can be classified into three major themes: “microservices development and operations in DevOps”, “approaches and tool support for MSA based systems in DevOps”, and “MSA migration experiences in DevOps”.

(3) We identified 24 problems with their solutions regarding implementing MSA in DevOps. These problems and solutions are classified into 8 categories (see Figure 6). We also identified three open research challenges that require further investigation (see Table 8).

(4) Concerning MSA description methods, most of the studies describe MSA design by using boxes and lines, UML diagrams, and formal methods. Moreover, three ADLs including CAOPLE, SDA, and Jolie were used or proposed for describing MSA in DevOps context.

(5) We identified 38 MSA patterns reported in the selected studies, but many of them have been not applied frequently. The most common MSA patterns reported in the studies are Circuit Breaker (5/38) and Migration (4/38) patterns.

(6) Regarding QAs when employing MSA in DevOps, we identified 15 QAs that are positively and negatively affected. The results show that most of the QAs are positively affected and certain QAs are affected both positively and negatively.

(7) We identified 50 tools that support building MSA based systems in DevOps, and further classified these tools into seven categories according to their functionalities (see Figure 7). GitHub as a version control system, Jenkins as a continuous integration server, and Puppet as a configuration management system, are the most popular tools.
(8) We observed that the combination of MSA and DevOps has been applied in a wide range of application domains, in which "Software Development Tools and Framework" and "Telecommunication" have received the most attention.

The findings of this SMS will benefit researchers who are interested in understanding the state of research of MSA in DevOps and conducting further research to fill the open research issues discussed in Section 5.2. Moreover, the findings of this SMS will facilitate knowledge transfer to practitioners about the problems, solutions, challenges, MSA description methods, MSA patterns, and tools for implementing MSA in DevOps. We argue that practitioners need to bring more dedicated solutions for addressing the monitoring, security, and performance degradation issues of MSA based systems in DevOps. Finally, as the next step of our work, we plan to conduct a Systematic Grey Literature Review (SGLR) to identify the gap between research and practice regarding MSA in DevOps.

# Acknowledgement

This work is partially sponsored by the National Key R&D Program of China with Grant No. 2018YFB1402800.

# Appendix A. Selected Studies

Table 15. List of the Selected Studies of this SMS

| ID | Authors, Publication Title, and Venue | Citation Count | Quality Score | Citation |
|---|---|---|---|---|
| S01 | Claus Pahl, Pooyan Jamshidi, and Olaf Zimmermann. **Architectural Principles for Cloud Software**. ACM Transactions on Internet Technology, 11(2): 1-23, 2018. | 31 | 3.1 | [61] |
| S02 | Robert Heinrich, André van Hoorn, Holger Knoche, Fei Li, Lucy Ellen Lwakatare, Claus Pahl, Stefan Schulte, and Johannes Wettinger. **Performance Engineering for Microservices: Research Challenges and Directions**. In: Proceedings of the 8th on International Conference on Performance Engineering Companion (ICPE), L'Aquila, Italy, pp. 223-226, ACM, 2017 | 52 | 3.2 | [62] |
| S03 | Mark Stillwell, Jose Coutinho. **A DevOps Approach to Integration of Software Components in an EU Research Project**. In: Proceedings of the 1st International Workshop on Quality-Aware DevOps (IWQAD), New York, USA, pp. 1-6, ACM, 2015. | 27 | 2.9 | [63] |
| S04 | Rion Dooley, Steven Brandt, and John Fonner. **The Agave Platform: An Open, Science-as-a-Service Platform for Digital Science**. In: Proceedings of the 2nd International Conferences on Practice and Experience on Advanced Research Computing (PEARC), Pittsburgh, USA, pp. 1-8, ACM, 2018. | 8 | 2.6 | [64] |
| S05 | Marco Miglierina, and Damian Tamburri. **Towards Omnia: A Monitoring Factory for Quality-Aware DevOps**. In: Proceedings of the 8th International Conference on Performance Engineering Companion (ICPE), L'Aquila, Italy, pp. 145-150, ACM, 2017 | 4 | 3.5 | [65] |
| S06 | Jihun Ha, Jungyong Kim, Heewon Park, Jaehong Lee, Hyuna Jo, Heejung Kim, and Jaeheon Jang. **A Web-Based Service Deployment Method to Edge Devices in Smart Factory Exploiting Docker**. In: Proceedings of the 8th International Conference on Information and Communication Technology Convergence (ICTC), Jeju, South Korea, pp. 708-710, IEEE, 2017. | 7 | 2.5 | [66] |
| S07 | Mei Chen, Rick Kazman, Serge Haziyev, Valentyn Kropov, and Dmitri Chtchourov. **Architectural Support for DevOps in a Neo-Metropolis BDaaS Platform**. In: Proceedings of the 34th Symposium on Reliable Distributed Systems Workshop (SRDSW), Montreal, Canada, pp. 25-30. 2015 | 7 | 3.3 | [67] |
| S08 | Desheng Liu, Hong Zhu, Chengzhi Xu, Ian Bayley, David Lightfoot, Mark Green, and Peter Marshall. **CIDE: An Integrated Development Environment for Microservices.** In: Proceeding of the 2nd International Conference on Services Computing (SCC), San Francisco, USA, pp. 808-812, IEEE, 2016 | 17 | 2.9 | [68] |
| S09 | Hui Kang, Michael Le, and Shu Tao. **Container and Microservice Driven Design for Cloud Infrastructure DevOps.** In: Proceeding of the 4th International Conference on Cloud Engineering (IC2E), Berlin, Germany, pp. 202-211, IEEE 2016. | 108 | 3.6 | [69] |
| S10 | Tran Thanh Quang, Stefan Covaci, Thomas Magedanz, Panagiotis Gouvas, and Anastasios Zafeiropoulos. **Embedding Security and Privacy into the Development and Operation of Cloud Applications and Services**. In: Proceedings of the | 12 | 3.1 | [70] |
| | 17th International Conference on Telecommunications Network Strategy and Planning Symposium (NETWORKS), Montreal, Canada, pp. 31-36, IEEE, 2016. | | | |
| S11 | Hong Zhu and Ian Bayley. **If Docker is the Answer, what is the Question?** In: Proceedings of the IEEE Symposium on Service-Oriented System Engineering (SOSE), Bamberg, Germany, pp. 152-163, 2018. | 3 | 3.6 | [71] |
| S12 | Armin Balalaie, Abbas Heydarnoori, and Pooyan Jamshidi. **Microservices Architecture Enables DevOps: Migration to A Cloud-Native Architecture**. IEEE Software, 33(3): 42-52, IEEE, 2016. | 338 | 4 | [2] |
| S13 | Yuan Fan, and Shang-Pin Ma. **Migrating Monolithic Mobile Application to Microservice Architecture: An Experiment Report.** In: Proceeding of 13th International Conference on AI & Mobile Services (AIMS), Honolulu, USA, pp. 109-112, IEEE, 2017. | 15 | 3.3 | [72] |
| S14 | Jeongju Bae, Chorwon Kim, and JongWon Kim. **Automated Deployment of Smartx IOT-Cloud Services Based on Continuous Integration**. In: Proceedings of 8th International Conference on Information and Communication Technology Convergence (ICTC), Jeju, South Korea, pp. 1076-1081, IEEE, 2016. | 3 | 2.7 | [73] |
| S15 | Cornel Barna, Hamzeh Khazaei, Marios Fokaefs, and Marin Litoiu. **Delivering Elastic Containerized Cloud Applications to Enable DevOps.** In: Proceedings of the 12th International Symposium on Software Engineering for Adaptive and Self-Managing Systems (SSASMS), Buenos Aires, Argentina, pp. 65-75, IEEE, 2017. | 21 | 2.8 | [74] |
| S16 | Hong Chen-Mei, Rick Kazman, Serge Haziyev, Valentyn Kropov, and Dmitri Chtchourov. **Big Data As A Service: A Neo-Metropolis Model Approach for Innovation**. In: Proceedings of the 49th Hawaii International Conference on System Sciences (HICSS), Koloa, HI, USA, pp. 5457-5466, IEEE, 2016. | 2 | 3.4 | [75] |
| S17 | Paul Drews, Ingrid Schirmer, Bettina Horlach, and Carsten Tekaat. **Bimodal Enterprise Architecture Management: The Emergence of a New EAM Function for a BizDevOps-Based Fast IT**. In: Proceedings of 21st International Enterprise Distributed Object Computing Workshop (EDOCW), Quebec City, Canada, pp. 57-64, IEEE, 2017 | 9 | 2.8 | [76] |
| S18 | George Pallis, Demetris Trihinas, Athanasios Tryfonos, and Marios Dikaiakos. **DevOps as a Service: Pushing the Boundaries of Microservice Adoption**. IEEE Internet Computing, 22(3): 65-71, IEEE, 2018 | 9 | 3 | [77] |
| S19 | Thomas Soenen, Steven Van Rossem, Wouter Tavernier, Felipe Vicens, Dario Valocchi, Panos Trakadas, Panos Karkazis et al. **Insights from SONATA: Implementing and integrating a Microservice-Based NFV Service Platform with a DevOps Methodology**. In: Proceedings of 30th IEEE/IFIP Network Operations and Management Symposium (NOMS), Taipei, Taiwan, pp. 1-6, IEEE, 2018 | 5 | 3.2 | [78] |
| S20 | Lianping Chen. **Microservices: Architecting for Continuous Delivery and DevOps**. In: Proceedings of 2nd International Conference on Software Architecture (ICSA), Seattle, USA, pp. 39-46, IEEE, 2018. | 44 | 3.5 | [79] |
| S21 | Tianlei Zheng, Xi Zheng, Yuqun Zhang, Yao Deng, ErXi Dong, Rui Zhang, and Xiao Liu. **SmartVM: A SLA-Aware Microservice Deployment Framework**. World Wide Web, 21(3): 1-19, Springer, 2018 | 6 | 2.9 | [80] |
| S22 | Pethuru Raj, Anupama Raman. **Automated Multi-Cloud Operations and Container Orchestration.** In: Software-Defined Cloud Centres: Operational and Management Technologies and Tools, pp. 185-218.Springer, 2018. | 3 | 2.8 | [81] |
| S23 | Bob Familiar, Jeff Barnes. **DevOps Using PowerShell, ARM, and VSTS**. In: Business in Real-Time Using Azure IoT and Cortana Intelligence Suite: Driving Your Digital Transformation, pp. 21-93, Springer, 2017. | 0 | 2.8 | [82] |
| S24 | Miika Kalske, Niko Mäkitalo, and Tommi Mikkonen. **Challenges When Moving from Monolith to Microservice Architecture**. In: Proceedings of 13th International Conference on Web Engineering (ICWE), Koloa, HI, USA, pp. 32-47 Springer, 2018 | 17 | 2.5 | [83] |
| S25 | Maurizio Gabbrielli, Saverio Giallorenzo, Claudio Guidi, Jacopo Mauro, and Fabrizio Montesi. **Self-Reconfiguring Microservices**. In: Theory and Practice of Formal Methods, pp. 194-210, Springer, 2015. | 27 | 3.1 | [84] |
| S26 | Bob Familiar. **From Monolithic to Microservice**. In: Microservices, IoT, and Azure: Leveraging DevOps and Microservice Architecture to Deliver SaaS Solutions, Chapter 1, pp. 1-7, Springer, 2015. | 1 | 2.6 | [85] |
| S27 | Armin Balalaie, Abbas Heydarnoori, and Pooyan Jamshidi. **Migrating to Cloud-Native Architectures Using Microservices: An Experience Report**. In: Proceedings of the 4th European Conference on Service-Oriented and Cloud Computing (ESOCC), Taormina, Italy, pp. 201-215, Springer, 2015. | 134 | 3.5 | [86] |
| S28 | Bob Familiar. **What is a Microservice**? Microservices, IoT, and Azure: Leveraging DevOps and Microservice Architecture to Deliver SaaS Solutions, Chapter 2, pp. 9-19, Springer, 2015. | 2 | 2.6 | [87] |
| S29 | Andreas Christoforou, Martin Garriga, Andreas S. Andreou, and Luciano Baresi. **Supporting the Decision of Migrating to Microservices through Multi-Layer Fuzzy Cognitive Maps**. In: Proceedings of the 14th International Conference on Service-Oriented Computing (ICSOC), Malaga, Spain, pp.471-480, Springer, 2017. | 6 | 3 | [88] |
| S30 | Maurizio Cavallari, Di Francesco Tornieri, and Marco De Marco. **Organizational Impact on Software Development of eServices Techniques**. In: Proceedings of the 8th International Conference on Exploring Services Science (IESS), Rome, Italy, pp. 64-75, Springer, 2017. | 1 | 2.1 | [89] |
| S31 | Cloves Carneiro Jr and Tim Schmelmer. **Deploying and Running Microservices.** In: Microservices From Day One: Build Robust and Scalable Software from the Start, pp. 151-174, Springer, 2016. | 14 | 2.6 | [90] |
| S32 | Martinez Manuel Perez, Tímea László, Norbert Pataki, Csaba Rotter, and Csaba Szalai. **Multivendor Deployment Integration for Future Mobile Networks**. In: Proceedings of International Conference on Current Trends in Theory and Practice of Informatics (SOFSEM), Krems, Austria, pp. 351-364, Springer, 2018. | 1 | 2.8 | [91] |
| S33 | Zykov Sergey, **Agile Services**. In: Managing Software Crisis: A Smart Way to Enterprise Agility, pp. 65-105, Springer, 2018 | 2 | 2.9 | [92] |
| S34 | Chorwon Kim, Seungryong Kim, and JongWon Kim. **Understanding Automated Continuous Integration for Containerized Smart Energy IoT-Cloud Service**. In: Proceedings of the 2nd International Conference on Ubiquitous Information Technologies and Applications (CUTE), Taichung, Taiwan, pp. 1275-1280, Springer, 2017 | 1 | 3.2 | [93] |
| S35 | Giuliano Casale, Cristina Chesta, Peter Deussen, Elisabetta Di Nitto, Panagiotis Gouvas, Sotiris Koussouris, Vlado Stankovski Andreas Symeonidis, Vlassis Vlassiou, Anastasios Zafeiropoulos, and Zhiming Zhao. **Current and Future Challenges of Software Engineering for Services and Applications**. Procedia Computer Science, 98: 34-42, Science Direct, 2016 | 18 | 2.6 | [94] |
| S36 | Kai Jander, Lars Braubach, and Alexander Pokahr**. Defence-in-depth and Role Authentication for Microservice Systems.** Procedia Computer Science, 130: 456-463, Science Direct, 2018. | 4 | 2.9 | [95] |
| S37 | Christof Ebert, Gallardo, Gorka, Hernantes, Josune, Serrano, and Nicola. **DevOps**. IEEE Software, 33(3): 94-100, IEEE, 2016. | 141 | 3.4 | [96] |
| S38 | Liming Zhu, Len Bass, and George Champlin-Scharff. **DevOps and its Practices**. IEEE Software, 33(3): 32-34, IEEE, 2016 | 65 | 2.6 | [97] |
| S39 | Tobias Schneider, and A. Wolfsmantel. **Achieving Cloud Scalability with Microservices and DevOps in the Connected Car Domain.** In: Proceedings of 1st Central Europe Workshop on Continuous Software Engineering (CEUR-WS), pp. 138-141, 2016 | 13 | 2.7 | [98] |
| S40 | Stefan Kehrer and Wolfgang Blochinger. **AUTOGENIC: Automated Generation of Self-configuring Microservices**. In: Proceedings of the 8th International Conference on Cloud Computing and Services Science (CLOSER), Funchal, Madeira, Portugal, pp. 35-46, Springer, 2018. | 9 | 3.5 | [99] |
| S41 | Sharma Sanjeev. **DevOps Plays for Driving Innovation**. In: The DevOps Adoption Playbook: A Guide to Adopting DevOps in a Multi-Speed IT Enterprise, pp. 189-260, Wiley, 2017. | 43 | 3.2 | [100] |
| S42 | Erich, Chintan Amrit, and Maya Daneva. **A Qualitative Study of DevOps Usage in Practice**. Journal of Software: Evolution and Process 29(6): 1-20, Wiley, 2017. | 42 | 3.3 | [101] |
| S43 | Karl, Holger, Sevil Dräxler, Manuel Peuster, Alex Galis, Michael Bredel, Aurora Ramos, Josep Martrat, Muhammad Shuaib Siddiqui, Steven van Rossem, Wouter Tavernier, George Xilouris. **DevOps for Network Function Virtualisation: An Architectural Approach**. Transactions on Emerging Telecommunications Technologies, 27(9): 1206-1215, Wiley, 2018. | 37 | 3.1 | [102] |
| S44 | Rory O'Connor, Peter Elger, and Paul M. Clarke. **Continuous Software Engineering - A Microservices Architecture Perspective**. Journal of Software: Evolution and Process, 29(11): 1-12, Wiley, 2017. | 29 | 3 | [103] |
| S45 | Daniel Russo, Vincenzo Lomonaco, and Paolo Ciancarini. **A Machine Learning Approach for Continuous Development**. In: Proceedings of the 5th International Conference on Software Engineering for Defence Applications (SEDA), Rome, Italy, pp. 109-119, Springer, 2018. | 3 | 3.2 | [104] |
| S46 | Prieto Zúñiga, Miguel, Emilio Insfran, Silvia Abrahão, and Carlos Cano-Genoves. **Automation of The Incremental Integration of Microservices Architectures**. In: Proceedings of the 25th International Conference on Information Systems Development (ISD), Katowice, Poland, pp. 51-68, Springer, 2017. | 2 | 3.2 | [105] |
| S47 | Davide Taibi, Valentina Lenarduzzi, and Claus Pahl. **Processes, Motivations, and Issues for Migrating to Microservices Architectures: An Empirical Investigation**. IEEE Cloud Computing, 4(5): 22-32, IEEE, 2017. | 73 | 3.2 | [106] |